\documentclass[preprint,amsmath,amssymb,aps,tikz,border=2mm,]{revtex4-1}
\usepackage{graphicx}
\usepackage{epsfig}
\usepackage{tabularx}
\usepackage{booktabs}
\usepackage{multirow}
\usepackage{amssymb}
\usepackage{amsmath}
\usepackage{standalone}
\usepackage{float}
\usepackage{makeidx}
\floatstyle{boxed}
\usepackage{caption}
\usepackage{subcaption}
\RequirePackage[colorlinks,citecolor=blue,urlcolor=blue,linkcolor=blue]{hyperref}
\usepackage{dcolumn}
\usepackage{pgfplots}
\usepgfplotslibrary{units}
\usepackage{mathtools}
\usepackage{filecontents}
\begin{document} 
\preprint{PRD}

\title{
Stability of neutrino parameters and self-complementarity relation with varying SUSY breaking scale}
\author{K. Sashikanta Singh\footnote{Corresponding author: ksm1skynet$@$gmail.com}}
\author{Subhankar Roy\footnote{meetsubhankar$@$gmail.com}}%
\affiliation{%
 Physics Department, Gauhati University, Guwahati 781014, India
}

\author{N. Nimai Singh\footnote{nimai03$@$yahoo.com}}
\affiliation{
 Physics Department, Manipur University, Imphal 795003\\
}
\date{\today}
             

\begin{abstract}

The scale at which supersymmetry (SUSY) breaks ($m_{s}$) is still unknown. The present article, following a top-down approach, endeavors to study the effect of varying $m_s$ on the radiative stability of the observational parameters associated with the neutrino mixing. These parameters get additional contributions in the minimal supersymmetric model\,(MSSM). A variation in $m_s$ will influence the bounds for which the Standard Model\,(SM) and MSSM work and hence will account for the different radiative contributions received from both sectors respectively, while running the  renormalization group  equations\,(RGE). The present work establishes the invariance of the self complementarity relation among the three mixing angles, $\theta_{13}+\theta_{12}\approx\theta_{23}$ against the radiative evolution. A similar result concerning the mass ratio, $m_2: m_1$ is also found to be valid. In addition to varying $m_s$, the work incorporates a range of different seesaw\,(SS) scales and tries to see how the latter affects the parameters.

\end{abstract}


\maketitle

\section{Introduction}

The physics of neutrino is going through a revolutionary period. From various recent experiments, a small but nonzero value of the reactor angle, $\theta_{13}$ is confirmed\cite{deSalas:2017kay,Gonzalez-Garcia:2015qrr}. In addition to this, the Dirac CP phase, $\delta$ is also observed \cite{Intro:2011zz,Gonzalez-Garcia:2013jma}. Recent experiments on neutrino oscillation, $0\nu\beta\beta$, and the cosmological observations have revealed precise and important results on the observational parameters like the three mixing angles\,($\theta_{13}$, $\theta_{12}$,\,$\theta_{23}$ ), two mass-squared differences ($\Delta m_{21}^2$,$\Delta m_{31}^2$) and possible upper bound on the sum of neutrino masses ($\Sigma m_{i}$) etc. \cite{bound2:2015ixa,bound3:2014jca,bound5:2014qwa}. But still we are unable to understand the absolute value of neutrino masses, nature of neutrino mass hierarchy, or its type: Dirac/Majorana etc. The realization that neutrinos are massive in contrast to its old popular assumption that it is massless (according to the SM) is one of the strong signatures that the SM of particle physics has to be extended beyond its present horizon.

Most of the current studies on physics beyond the SM\,(BSM) relies on the possible existence of supersymmetry (SUSY). But there are other models of BSM physics which does not incorporate the idea of SUSY \cite{Melfo:2007sr,Dorsner:2005fq}. It is hypothesized that SUSY existed at the early stage of big bang. But with the expansion of our Universe SUSY gets broken and reduced to our present day SM. At what scale that breaking occurs is still an unknown but an important parameter. The general idea is that there are two possible energy scales for the SUSY breaking $(m_s)$: low and high. The low $m_s$ scale \cite{Dudas:2012fa,Dine:1996xt} is expected to be about a few TeV or so as suggested by the grand unified theory (GUT), whereas the high SUSY breaking scale is expected to be somewhere around $10^{12}$ GeV \cite{Feldstein:2012bu}.

One significant finding from the recent LHC experiment which sounds a little disappointing towards the possibility of SUSY is that the experiment, which was operated at an energy scale of $13$ TeV, has not provided any evidence of the existence of SUSY particles so far \cite{Aaij:2015bfa,Kazana:2016gni}. In SUSY inspired neutrino physics, it is predicted that SUSY plays an important role over the neutrino masses and other observational parameters \cite{Bustamante:2010bf,Cadiz:2013yja,Mukhopadhyaya:2003te}. The gauge coupling and Yukawa coupling constants suffer different radiative contributions from the MSSM and SM sectors. Similar to this, we expect that the neutrino observational parameters are also subjected to such kind of effects. 

One of the reasons why the variation in $m_s$ is expected to bring changes to various observational parameters is owing to the changes in the effective range of both MSSM and SM. When we increase the $m_s$ scale, the effective range of SM increases, whereas that for MSSM decreases and vice versa. It will change the amount of radiative correction that each parameter receives from the SM and MSSM, respectively. In our previous work \cite{Singh:2015kua}, we show the variation of the unification point of the gauge couplings  with varying $m_s$ scale. Such behavior is likely to be seen for the neutrino oscillation parameters too. In this regard, it is important to study the possible effects of varying $m_s$ on the radiative evolution of the neutrinos and hence ,to determine (or narrow down) the possible range of $m_s$ scale.


The possible reason behind the suppression of SUSY motivated effects at the LHC experiments may be due to the low luminosity of the beam. By the end of 2012 LHC's integrated luminosity, running at a center-of-mass energy $\sqrt{s}= 8$TeV, is already over $20fb^{-1}$ \cite{Apollinari:2017cqg}. The present integrated luminosity of the LHC for $\sqrt{s}= 13$TeV is $35.9fb^{-1}$ for CMS \cite{CMS-PAS-B2G-17-002} and $36.1fb^{-1}$ for ATLAS \cite{Aaboud:2017fgj}. Some predicted the required integrated luminosity for observing SUSY related events to be $3000\,fb^{-1}$ \cite{Gladyshev:2012xq,Demidov:2013aka}, which is approximately 85 times  greater than the present luminosity. Nevertheless, this still gives us a hope for the possible existence of $m_s < 13\,TeV$.  If seesaw (SS) mechanism is the only cause behind the generation of small neutrino masses, then it appears that the right handed neutrino mass scale must lie somewhere within the  range of $(10^{10}-10^{16})$GeV \cite{Ren:2017xyv,Arbelaez:2011bb}. In our analysis we shall vary the SS scale starting from $10^{10}$-$10^{15}$\,GeV.

One sees that the numerical range of three mixing angles within $1\sigma$\cite{deSalas:2017kay} appears as in the following:
\begin{eqnarray}
\theta_{13}= {8.44^0}^{+0.16}_{-0.17}, \,\theta_{12}={34.5^{0}}^{+1.1}_{1.0}\,\,\text{and}\,\,\theta_{23}={41.0^{0}}^{+1.1}_{-1.1}.
\end{eqnarray}
We see that there may lie a self-complementarity\,(SC) among these parameters in terms of the following relation
\begin{eqnarray}
\theta_{23}=q\times(\theta_{13}+\theta_{12}),\label{comp2}
\end{eqnarray} 

where, the parameter, $q$ is either unity or $\mathcal{O}(1)$. The SC is an important phenomenological relation \cite{Zhang:2012xu,Haba:2012qx} similar to the quark-lepton complementarity relations \cite{Minakata:2004xt,Minakata:2005rf,Zhang:2012zh}. The possible existence of such relations among the parameters are expected to be the signatures of a certain flavor symmetry working in the background. The present analysis attempts not to deal with the possible origin of such a kind of an SC relation rather, it insists on the existence of such a relation even at higher energy scale. Our work starts with an assumption that this SC relation holds good at the SS scale. Through our analysis, we will show that this relation remains invariant against the radiative evolution for varying the $m_s$ and SS scale. We emphasize that similar to the works in the literature which focus only on the renormalization group invariant parameters \cite{Haba:2013kga,Foot:2007ay,Tsuyuki:2015kqa,Harrison:2010mt,Demir:2004aq}, the SC relation can also serve as an RGE invariant relation. 

The present investigation is a continuation of our previous work \cite{Singh:2015kua}, where we studied the radiative evolution of the three gauge, third generation Yukawa and quartic Higgs couplings following a bottom-up approach, with varying the SUSY breaking scale $m_s$. It was observed that the unification scales for both the gauge couplings and Yukawa couplings vary but in the opposite trend and tend to attain a fixed value with increasing $m_s$. There, we vary $m_s$ starting from 500\,GeV to 7\,TeV. However, in the present work, we follow the top down approach starting from the seesaw scale up to the electroweak scale. We fix, $\tan\beta=58.6$, which is relevant in the context of our previous work \cite{Singh:2015kua}.

This paper is organized in the following order. In Sec. II, we give a brief discussion of the neutrinos RGEs. In Sec. III, we study the possible radiative effects on the neutrino parameters at the weak scale. In Sec. IV, we present the numerical analysis. In Sec. V, we  summarize our work, and in the Appendix, we give the RGEs for the gauge, Yukawa and quartic Higgs couplings in two loops for both the SM and MSSM.

 

\section{RGEs for neutrino parameters}
Renormalization group approach is a tool for studying physics at a different energy scale, which are otherwise impossible to reach with the current technology, and then to compare it with the available low energy data. Radiative analysis of neutrino parameters requires the RGEs of gauge couplings, Yukawa couplings and the quartic Higgs couplings. The radiative properties of these couplings have been studied extensively in different models and these three gauge couplings are expected to be unified at an energy scale approximately at $2\times 10^{16}GeV$ \cite{gy1:1992ac,gy2:1998dh,gy3:1994qm,gy4:1997wx,gy5:1992kv,n5:2003kp}. The RGEs for the gauge couplings, Yukawa couplings, and quartic Higgs coupling are given in the Appendix. We use 2-loops RGEs for both the SM and MSSM.

The RGE analysis of the neutrino parameters can be done in two possible ways viz: i) by a run and diagonalize method: where the whole neutrino mass matrix is allowed to evolve using their appropriate RGEs and then the corresponding neutrino parameters can be achieved at the desirable energy scale ($\mu$) by diagonalizing the neutrino mass matrix, ii) by using the RGEs of the corresponding neutrino parameters separately as defined by the equations.(\ref{t12}) to (\ref{m3}). In both the cases, the RGEs of all the neutrino parameters and the RGEs  of various coupling parameters are required to be solved simultaneously.  In this work, we adopt the later stand. 

The input parameters for the gauge, Yukawa, and quartic Higgs couplings at the seesaw scale, given in Table \ref{gyl}, are taken form our previous work \cite{Singh:2015kua}.  In the present analysis, we choose our starting energy scale to be the SS scale.  We consider a different possible SS scale starting from $10^{10}$GeV to $10^{15}GeV$, and  we run down all the observational neutrino parameters from SS scale up to the electroweak scale ($m_Z=91.18 GeV$) via $m_s$, which also varies in our analysis.

The radiative properties of neutrinos has been studied extensively in various models \cite{n5:2003kp,n1:1999tg,n3:2014hya,n4:1993tx,n6:2000ma,n7:2006ye,n8:2014uza,n9:2008wn,n10:2014lwa,a3:2005dr}. The standard two loops RGEs for the neutrino masses, mixings, and CP phases are shown below. For the three neutrino mixing angles \cite{n5:2003kp}, the RGEs are

\begin{eqnarray}
\dot{\theta}_{12} &=& -\frac{Cy_{\tau}^2}{32\pi^2} \sin 2\theta_{12} s^2_{23}\frac{|m_1e^{i\psi_1}+m_2e^{i\psi_2}|^2}{\Delta m^2_{21}},	\label{t12} \\
\dot{\theta}_{13} &=& -\frac{Cy_{\tau}^2}{32\pi^2} \sin 2 \sin 2\theta_{23} \frac{m_3}{\Delta m^2_{31}{(1+\xi)}}	\label{t13} \\ 
 & & \times \left[m_1\cos\left(\psi_1-\delta\right)-\left(1+\xi\right)m_2\cos\left(\psi_2-\delta\right)-\xi m_3\cos\delta\right], \\ \nonumber
 \dot{\theta}_{23} &=& -\frac{Cy_{\tau}^2}{32\pi^2}\sin 2\theta_{23}\frac{1}{\Delta m_{31}^2}\left[c_{12}^2|m_2e^{i\psi_2}+m_3|^2+s^2_{12}\frac{|m_1e^{1\psi_2}+m_3|^2}{(1+\xi)}\right],\label{t23} 
\end{eqnarray}
where, $\Delta m^2_{21}=m_2^2-m_1^2$ and $\Delta m^2_{31}=m_3^2-m_1^2$, $\xi=\frac{\Delta m^2_{21}}{\Delta m^2_{31}}$.

\vspace{0.3cm}
The RGEs for the three phases are,

(for Dirac phase)
\begin{equation}
\dot{\delta} = \frac{Cy_{\tau}^2}{32\pi^2}\frac{\delta^{(-1)}}{\theta_{13}}+\frac{Cy_{\tau}^2}{8\pi^2}\delta^{0}, \label{dirac} 
\end{equation}
where, 
\begin{eqnarray}
\delta^{(-1)}&=&\sin 2\theta_{12}\sin 2\theta_{23}\frac{m_3}{\Delta m^2_{31}(1+\xi)}\times \\ \nonumber
& & \left[m_1\sin\left(\psi_1\delta\right)-(1+\xi)m_2\sin\left(\psi_2-\delta\right)+\xi m_3\sin\delta\right], \\
\delta^{(0)} &=& \frac{m_1m_2s^2_{23}\sin\left(\psi_1-\psi_2\right)}{\delta m^2_{21}} \\ \nonumber
& & +m_3s^2_{12}\left[\frac{m_1\cos 2\theta_{23}\sin\psi_1}{\Delta m^2_{31}(1+\xi)}+\frac{m_2c_{23}^2\sin\left(2\delta-\psi_2\right)}{\Delta m^2_{31}}\right] \\ \nonumber
& & +m_3c^2_{12}\left[\frac{m_1c^2_{23}\sin\left(2\delta-\psi_1\right)}{\Delta m^2_{31}(1+\xi)}+\frac{m_2\cos 2\theta_{23}^2\sin\left(\psi_2\right)}{\Delta m^2_{31}}\right],
\end{eqnarray}

(for Majorana phases)
\begin{eqnarray}
\dot{\psi}_1 &=& \frac{Cy_{\tau}^2}{8\pi^2}\left\{m_3\cos 2\theta_{23}\frac{m_1s^2_{12}\sin\psi_1+(1+\xi)m_2c^2_{12}\sin\psi_2}{\Delta m^2_{31}(1+\xi)}\right. \\ \nonumber
& & \left.+\frac{m_1m_2c_{12}^2s_{23}^2\sin\left(\psi_1-\psi_2\right)}{\Delta^2_{21}}\right\} ,\label{psi1} \\
\dot{\psi}_2 &=& \frac{Cy_{\tau}^2}{8\pi^2}\left\{m_3\cos 2\theta_{23}\frac{m_1s^2_{12}\sin\psi_1+(1+\xi)m_2c^2_{12}\sin\psi_2}{\Delta m^2_{31}(1+\xi)}\right. \\ \nonumber
& & \left.+\frac{m_1m_2s_{12}^2s_{23}^2\sin\left(\psi_1-\psi_2\right)}{\Delta^2_{21}}\right\} \label{psi2}. 
\end{eqnarray}

The RGEs for the neutrino mass eigenvalues are

\begin{eqnarray}
\dot{m_1} &=& \frac{1}{16\pi^2} \left[\alpha+Cy_{\tau}^2\left(2s_{12}^2s_{23}^2+F_1\right)\right]m_1, \label{m1} \\
\dot{m_2} &=& \frac{1}{16\pi^2} \left[\alpha+Cy_{\tau}^2\left(2c_{12}^2s_{23}^2+F_2\right)\right]m_2, \label{m2}\\
\dot{m_3} &=& \frac{1}{16\pi^2} \left[\alpha+2Cy_{\tau}^2c_{13}^2c_{23}\right]m_3, \label{m3}
\end{eqnarray}

where, 
\begin{eqnarray}
F_1 &=& -s_{13}\sin 2\theta_{12}\sin 2\theta_{23}\cos\delta +2s_{13}^2c^2_{12}c_{23}^2, \\ 
F_2 &=& s_{13}\sin 2\theta_{12}\sin 2\theta_{23}\cos\delta +2s_{13}^2s^2_{12}s_{23}^2,
\end{eqnarray}

\begin{equation}
\begin{rcases}
\alpha &= -\frac{6}{5}g_1^2-6g_2^2+6y_t^2 \\
  C &= 1
\end{rcases}
\text{ for MSSM}
\label{mssm}
\end{equation}

\begin{equation}
\begin{rcases}
\alpha &= -3g_2^2+2y_{\tau}^2+6y_t^2+6y_b^2 \\
  C &= \frac{1}{2}
\end{rcases}
\text{ for SM}
\label{sm}
\end{equation}

With all the necessary mathematical frameworks in hand, we can now study the radiative nature of neutrino masses, mixings and CP phases using the top-down running approach together with the MSSM unification conditions.

\begin{table}[t]
\centering
\begin{tabular}{cccc}
\hline 
 gauge coplings  & Yukawa couplings  & quartic-Higgs couplings \\ 
\hline
$g_1$ - 0.6032 & $y_t$ - 0.76809 & $\lambda$ - 0.58 \\
$g_2$ - 0.6826 & $y_b$ - 0.80488 & - \\
$g_3$ - 0.7557 & $y_{\tau}$ - 0.91448 & - \\
\hline 
\end{tabular} 
\caption{Input values for gauge, Yukawa  and quartic Higgs couplings\cite{Singh:2015kua}.}
\label{gyl}
\end{table}

In the first step all the parameters are allowed to run down from the seesaw scale to the SUSY breaking scale using their respective MSSM RGEs, and from the SUSY breaking scale further down to the electroweak scale using their SM RGEs. At the transition point from MSSM to SM, we apply appropriate matching conditions as shown below,

\begin{eqnarray}
g_i\left(M^-_{SUSY}\right) &=& g_i\left(M^+_{SUSY}\right), \\
\lambda_t\left(M^-_{SUSY}\right) &=& \lambda_t\left(M^+_{SUSY}\right)\sin \beta, \\
\lambda_b\left(M^-_{SUSY}\right) &=& \lambda_b\left(M^+_{SUSY}\right)\cos \beta ,\\
\lambda_{\tau}\left(M^-_{SUSY}\right) &=& \lambda_{\tau}\left(M^+_{SUSY}\right)\cos \beta,
\end{eqnarray}

where, $\displaystyle \tan\beta=v_u/v_d$ such that $\displaystyle v_u=v \sin\beta$, $\displaystyle v_d=v\cos\beta$ and $\displaystyle v=246 GeV$ is the vacuum expectation value (VEV) of the Higgs field. In our analysis, we choose a single SUSY spectrum for simplicity and study the radiative stability of the neutrino parameters at the weak scale for varying $m_s$.

\section{Radiative effects on the neutrino oscillation parameters and the CP phases}
The radiative effects on the neutrino parameters for a strict  normal  or inverted hierarchy is small. If the neutrinos masses have a quasidegenerate spectrum, then the RG evolution between the lowest seesaw scale and electroweak energy scale can have sizable effects \cite{Casas:1999ac,Casas:1999tg,Chen:2001gk,Joshipura:2002xa} on the neutrino oscillation parameters. The RG effects may even account for the difference between the mixings in the quark and the lepton sectors \cite{Mohapatra:2003tw}.


In MSSM, both the atmospheric $(\theta_{23})$ and solar mixing angle $(\theta_{12})$ increase with the decrease in energy as predicted by eq.(\ref{t12}) and eq.(\ref{t13}). Out of the three mixing angles, the solar mixing angle is prone to the largest RG effects because of the presence of small $\Delta m_{31}^2$ in the denominator, whereas $\theta_{13}$ is subjected to the smallest RG effect.


In the top-down approach, all the three mass eigenvalues behave in a similar fashion, and they all decrease with the decrease in energy scale. Because of the comparatively larger value of $\alpha$ with respect to $y_t,\ y_b,$ and $y_{\tau}$, the RG running effect on the mass eigenvalues is less. But, due to the same factor $\alpha$, there is appreciable running in the RGEs of the mass eigenvalues in the SM case. The running of the mass eigenvalues in the MSSM is defined by a common scaling factor, except for the case of a large $\tan\beta$ where it deviates considerably.

For nearly degenerate neutrino masses and a large $\tan\beta$, the radiative influence of CP phases over other parameters becomes important. All of the phases (both Majorana and Dirac) undergo radiative correction. For different sets of the input phases, the RG effects on the neutrino oscillation parameters may differ. In the context, when the two Majorana phases are equal\cite{n5:2003kp}, the evolutions of the parameters are highly suppressed since the leading terms in the RGEs of the phases become zero\,[See eq.(\ref{psi1})and eq. (\ref{psi2})].

\section{Numerical analysis and the Results}

The RGEs are differential equations and demand the input values for the parameters to be sought out, at the very outset. In our case, the starting point is the SS scale, and finally, we end up at the EW scale. From SS scale upto $m_s$ scale the RGEs follow  certain pattern (eq.(\ref{mssm})) and reverts to another form in the region from $m_s$ upto EW scale (eq.(\ref{sm})). Both the SS scale and $m_{s}$ are unknown to us. Our present analysis although tries to visualize the effect on the neutrino observational parameters for varying $m_s$, yet gives emphasis on the choice of the SS scale also. We fix the $m_{s}$ values in between 1\,TeV to 13\,TeV. In addition, the SS scale is also assigned certain fixed values between $10^{10}$ GeV to $10^{15}$\,GeV.    

\begin{figure}
  \begin{subfigure}[b]{0.4\textwidth}
    \includegraphics[width=\textwidth]{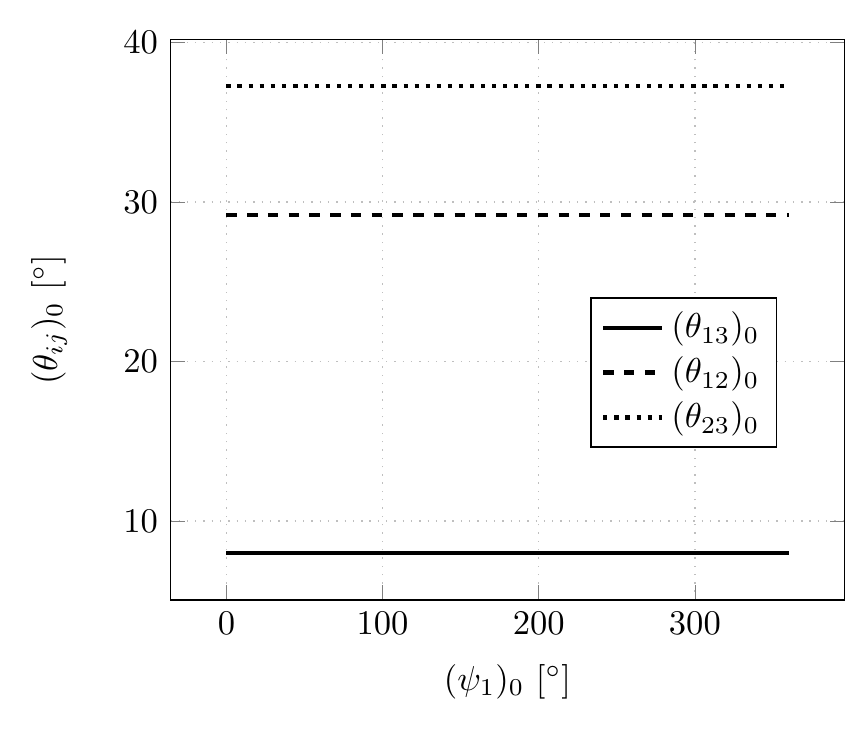}
    \subcaption{}
    \label{pa}
  \end{subfigure}
  \begin{subfigure}[b]{0.4\textwidth}
    \includegraphics[width=\textwidth]{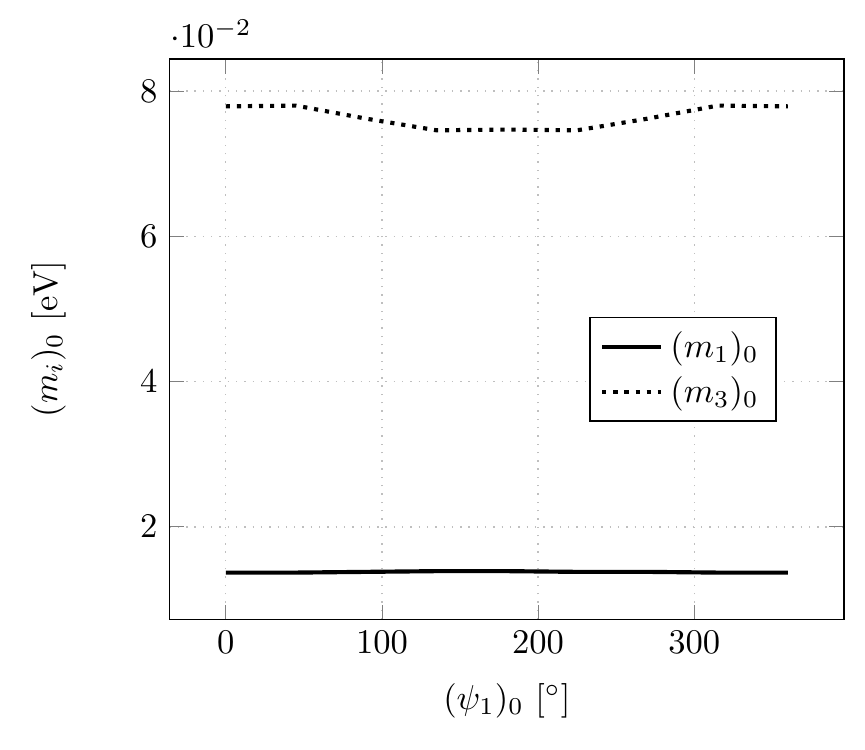}
    \subcaption{}
    \label{pm}
  \end{subfigure}
    \captionsetup{justification=raggedright, singlelinecheck=false,  width=0.95\linewidth}
  \caption{\footnotesize (a) The variation of the $(\theta_{ij})_0$ against $(\psi_i)_0$ is shown. (b)The stability of $(m_i)_0$ against $(\psi_i)_0$ is studied. In our calculations, we assume the Majorana parameters to be equal. The $m_s$ and SS scale are fixed at 5TeV and $10^{14}$ GeV, respectively. The other initial input, $(\delta)_0=90^{\circ}$ and $(m_2)_0= 2.34 \times 10^{-2}$GeV. The purpose of this study is to achieve the numerical values of the parameters at the EW scale within $1\sigma$.}
  \label{pam}
\end{figure}

The parameters, $g_1$, $g_2$, $g_3$, $y_{t}$, $y_{b}$, $y_{\tau}$ and $\lambda$ are specified as per Table.\ref{gyl}. In the present analysis, we have got nine free parameters: $m_1$, $m_2$, $m_3$, $\theta_{13}$, $\theta_{23}$, $\theta_{12}$, $\delta$, $\psi_{1}$ and $\psi_{2}$. As stated earlier, the present study presumes the existence of the SC relation (see eq.\,(\ref{comp2})) at the SS scale. By virtue of this relation, we assign initial input values only to $\theta_{13}$ and $\theta_{12}$. Further simplifications are made regarding the initial choice of $\psi_{1}$ and $\psi_{2}$, which are constrained to be equal,$(\psi_{1})_{0}=(\psi_{2})_{0}$, for all subsequent calculations (the notations $(...)_{0}$ represent the initial input value of the parameter within the bracket). In that way, we assign input values only to six neutrino observational parameters. To simplify, we summarize our strategy in the following way,

\begin{enumerate}
\item[(Step 1)] We vary the initial values of the six neutrino parameters at a fixed $m_s$ scale. To ensure that the initial choice of the parameters beget the numerical values at the EW scale which are consistent within $3\sigma$ range, we follow a simple mechanism. To illustrate, let us fix $m_{s}$ at $5\,TeV$, SS scale at $10^{14}\,GeV$, and assume, $(m_2)_{0}=2.34\times10^{-2}\,eV$ and $(\delta)_{0}=90^{\circ}$. The remaining parameters, $(\theta_{13})_{0}$, $(\theta_{23})_{0}$, $(m_1)_{0}$, $(m_3)_{0}$ and $(\psi_1)_{0}$ are assigned with certain numerical values, so that the final output at the EW scale lies within $3\sigma$. Next, we vary the parameter, $(\psi_{1})_{0}$ and see how the remaining parameters, like, $(\theta_{ij})_{0}$ and $(m_{i})_{0}$ are to be adjusted in order to keep the outcome within the $3\sigma$ range. For details, see Table \ref{vmp}, Figs.\,\ref{pa}, and \ref{pm}. We see that, except $(m_{3})_{0}$ which varies a little, the other input parameters are almost stable against changing $(\psi_{1})_0$. The motivation behind performing this step is to ensure that the final numerical values in concern with the neutrino observational parameters are not too  sensitive to the initial input of the Majorana phase. This observation helps us to choose an arbitrary value for $(\psi_1)_0$. We take $(\psi_{1})_{0}=45^{\circ}$ for all subsequent calculations.   
,
\item[(Step 2)] The SUSY breaking scale $m_{s}$, is attributed to the following numerical values like, $1,\,3,\,5,...13$\,TeV and in accordance with that, we categorize seven sets of input values as, $A1$, $A3$, $A5$...$A13$ respectively. For example, the set $A_{5}$ corresponds to the set of input $(\theta_{ij})_{0}$, $(m_{i})_{0}$, $(\delta)_0$, and $(\psi_{1})_0$, at $m_{s}=5\,TeV$. For all the above mentioned sets, we fix $(\delta)_0=90^{\circ}$. Similarly, we assign sets, $B1$, $B3$, $B5$...$B{13}$ with $(\delta)_0=270^{\circ}$. This is to be noted that both kinds of sets $Aj$ and $B{j}$ are the input values of the neutrino parameters, at the SS scale of $10^{14}$\,GeV. There is another $\mathcal{O}(1)$ parameter, $q$ which appears in eq.\,(\ref{comp2}) is tuned between $0.95$ to $0.97$. For details, see Table.\,\ref{90i}.

\item[(Step 3)] In this step, keeping a certain input set, say $A{5}$ fixed, we vary the $m_s$ scale between $1$\,TeV  to $13$\,TeV, and check the stability of the neutrino observational parameters at the EW scale. The details are shown in the Tables. ref{90m12} - \ref{90m3}.

\item[(Step 4)] We repeat step 3, for different values of the SS scale, such as $10^{10}$, $10^{11}$...$10^{15}$\,GeV.

\end{enumerate}

We will now discuss the results of our analysis.

\subsection{For varying $m_s$ at fixed SS scale:}

We keep track of the the numerical values of the neutrino observational parameters at the EW scale. From Tables.\, \ref{90m12}-\ref{90m3}, one sees that, except, $\Delta m_{21}^2$, other parameters like $\theta_{13}$, $\theta_{12}$, $\theta_{23}$, and $\Delta m_{31}^2$ show stability at the face of changing $m_{s}$. For all the three mixing angles, the fluctuations are consistent within $3\sigma$ bound\,\cite{deSalas:2017kay}.But for $\Delta m_{31}^2$, the fluctuations sometimes cross the $3\sigma$ bound. Although the input entries corresponding to different neutrino parameters are almost the same for all the sets $Aj$ and $Bj$, the solar mass squared difference at the EW scale is found quite sensitive towards both the initial input as well as to the $m_s$ scale. To illustrate, one can see that for the input data set, say $A5$, which results in $\Delta m_{21}^2=7.57\times 10^{-5}\,eV^2$, and this is consistent within $1\sigma$ bound, for $m_{s}$ being set at $5\,TeV$. If $m_{s}$ is changed a little, say to 3 TeV and 7 TeV, we see that for the same input data set $A5$, the $\Delta m_{21}^2$ become, $9.16\times 10^{-5}\,eV^2$ and $6.67\times 10^{-5}\,eV^2$ respectively. This output lies strictly outside the $3\sigma$ region. However, if we achieve an acceptable $\Delta m_{21}^2$, against a higher $m_s$ scale, we can expect a little stability. To exemplify, if for $A11$, we achieve, $\Delta m_{21}^2=7.54\times 10^{-5}\,eV^2$ (within $1\sigma$ bound), against $m_s=11\,TeV$, then changing the $m_s$ to either $9$ or $13$ TeV, will not take this parameter outside $3\sigma$. In addition, both solar and atmospheric mass squared difference decreases, with the increase in $m_s$ scale. The CP violating phases also vary a little if $m_{s}$ were changed. With the increase of the latter, $\delta$ decreases, whereas, the two Majorana phases, increase (See Fig.(\ref{dpmsss})).

\subsection{For varying $m_s$ and SS scale:}

The discussion concerned so far, is true only for the SS scale: $10^{14}$\,GeV. We try to see how a changing SS scale, along with $m_s$, can affect the physical parameters at the EW scale as per the step(4) mentioned above. We note down the following. To exemplify, let us choose the input data set $B5$, which is capable of producing, observable parameters at the EW scale consistent within $3\sigma$, with $m_{s}$ being fixed at $5$\,TeV, and the SS scale at $10^{14}$\,GeV. With the SS scale fixed, first we vary $m_{s}$ and we get certain plot, which shows how the numerical value of that observable parameter at EW scale changes against $m_s$. We redo the same to get another plot, but at a different SS scale, for same input data set. We observe the ascent or descent of the plots against the different SS scale.

\begin{figure}
  \begin{subfigure}[b]{0.4\textwidth}
    \includegraphics[width=\textwidth]{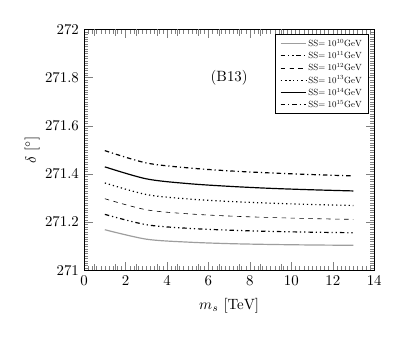}
\captionsetup[figure]{font=small,skip=0pt}  
  \subcaption{}
    \label{dmsss}
  \end{subfigure}
  \begin{subfigure}[b]{0.4\textwidth}
    \includegraphics[width=\textwidth]{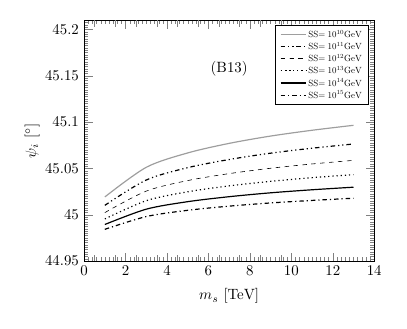}
\captionsetup[figure]{font=small,skip=0pt}
    \subcaption{}
    \label{pmsss}
  \end{subfigure}
  \caption{\scriptsize The fluctuations of (a) the Dirac phase ($\delta$) and (b) the Majorana phase ($\psi_i$) after RG evolution, at the EW scale, against changing $m_s$, and the SS scale are studied. $m_s$ values are fixed at 1 Tev, 3 TeV, 5 Tev, 7 TeV, 9 Tev, 11 Tev, 13 TeV, and different SS scales are assumed at $10^{10}$ GeV, $10^{11}$ GeV, $10^{12}$ GeV, $10^{13}$ GeV, $10^{14}$ GeV, and $10^{15}$ GeV. Here, we consider only one input data set B13 as in Table.\,(\ref{90i}).}
    \label{dpmsss}
\end{figure}

\begin{figure}
  %
  \begin{subfigure}[b]{0.4\textwidth}
    \includegraphics[width=\textwidth]{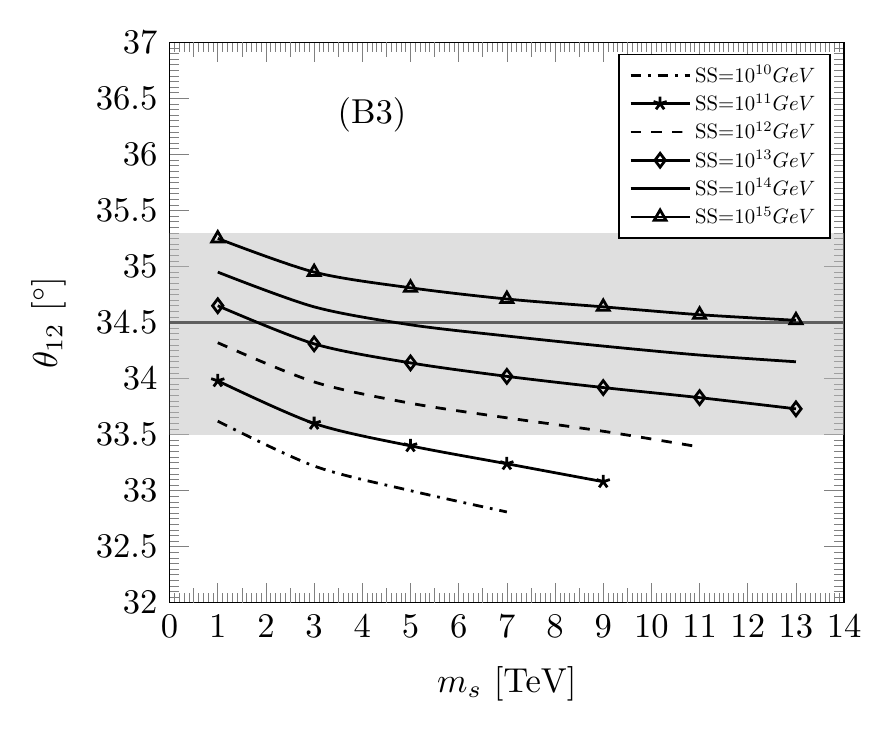}
\captionsetup[figure]{font=small,skip=0pt}
    \subcaption{}
    \label{t123}
  \end{subfigure}
\vspace*{1mm}  
    \begin{subfigure}[b]{0.4\textwidth}
    \includegraphics[width=\textwidth]{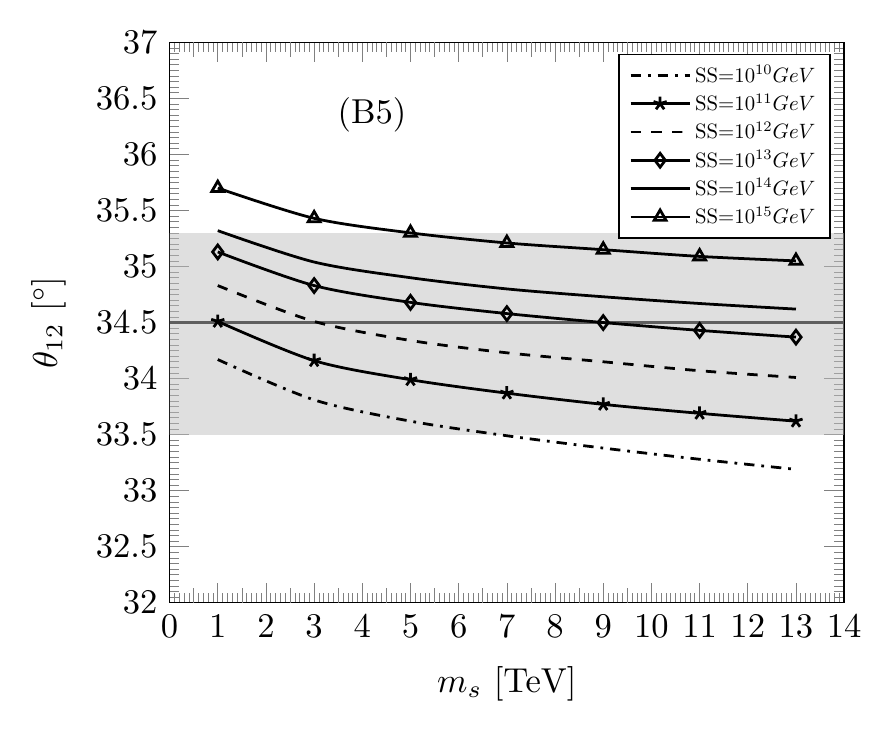}
    \subcaption{}
    \label{t125}
  \end{subfigure}
  \begin{subfigure}[b]{0.4\textwidth}
    \includegraphics[width=\textwidth]{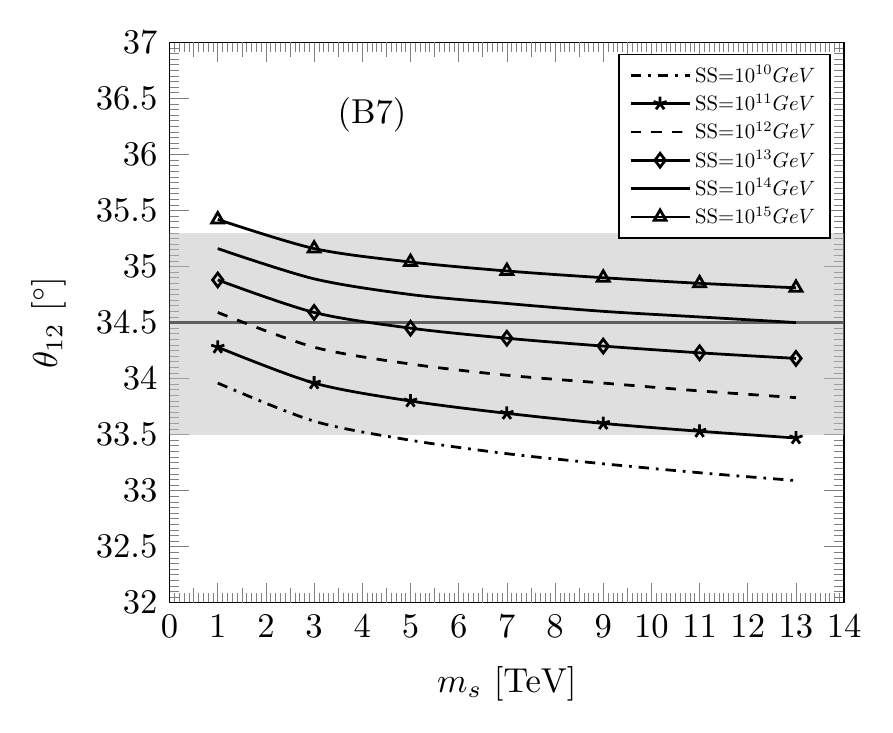}
    \subcaption{}
    \label{t127}
  \end{subfigure}
\vspace*{1mm}  
    \begin{subfigure}[b]{0.4\textwidth}
    \includegraphics[width=\textwidth]{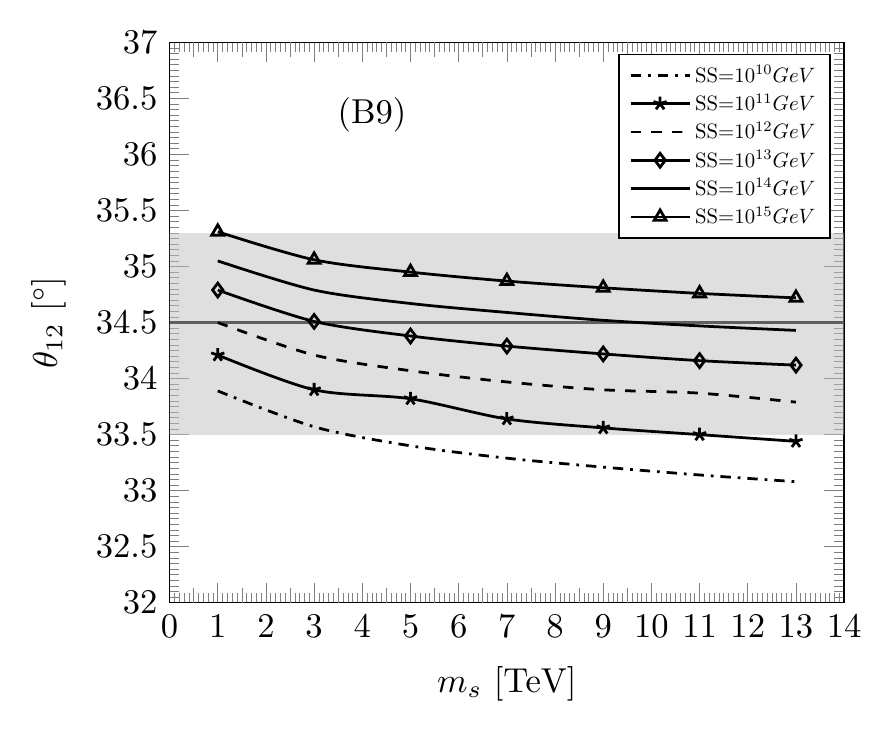}
    \subcaption{}
    \label{t129}
  \end{subfigure}
  \begin{subfigure}[b]{0.4\textwidth}
    \includegraphics[width=\textwidth]{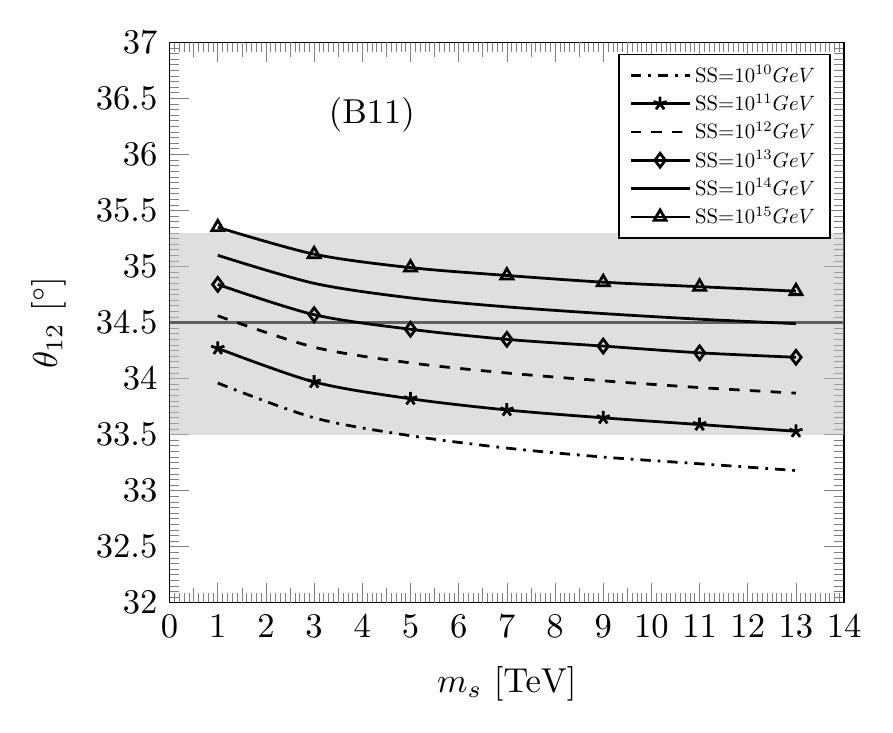}
    \subcaption{}
    \label{t1211}
  \end{subfigure}
    \begin{subfigure}[b]{0.4\textwidth}
    \includegraphics[width=\textwidth]{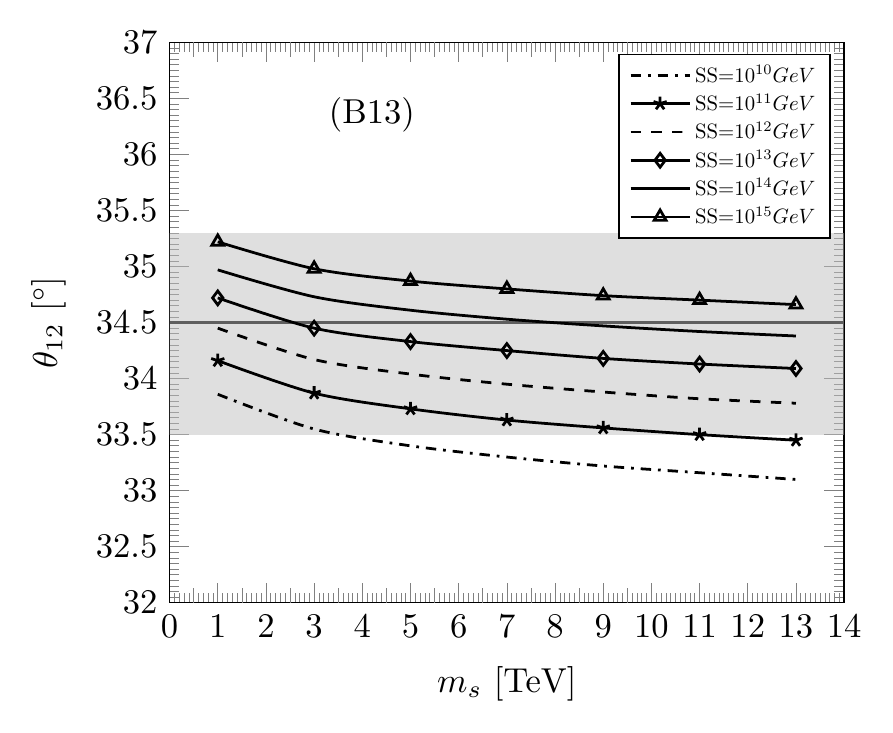}
    \subcaption{}
    \label{t1213}
  \end{subfigure}
 \caption{\scriptsize The fluctuations of the numerical values of $\theta_{12}$, at the EW scale is studied, against changing $m_s$ , and SS scale. The shaded region (horizontal) represents the experimental $3\sigma$ range \cite{deSalas:2017kay} and the horizontal bold line inside the shaded region indicates the best-fit value. The six figures (a), (b), (c), (d), (e), and (f) are for the different input data sets B3, B5, B7, B9, B11, and B13 respectively (as given in Table.\,(\ref{90i})). The SS scales are fixed at $10^{10}$ GeV, $10^{11}$ GeV, $10^{12}$ GeV, $10^{13}$ GeV, $10^{14}$ GeV, and $10^{15}$ GeV.}
  \label{t12msss}
\end{figure}

\begin{figure}
  %
  \begin{subfigure}[b]{0.4\textwidth}
    \includegraphics[width=\textwidth]{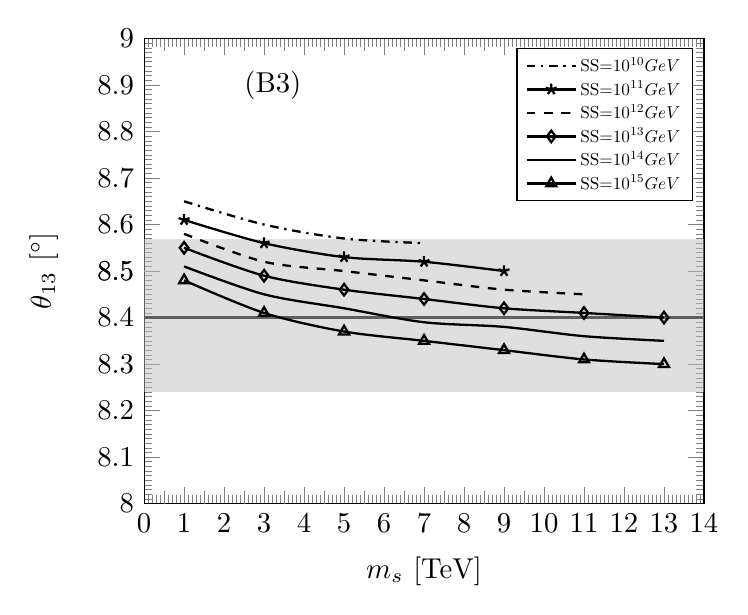}
\captionsetup[figure]{font=small,skip=0pt}
    \subcaption{}
    \label{t133}
  \end{subfigure}
    \begin{subfigure}[b]{0.4\textwidth}
    \includegraphics[width=\textwidth]{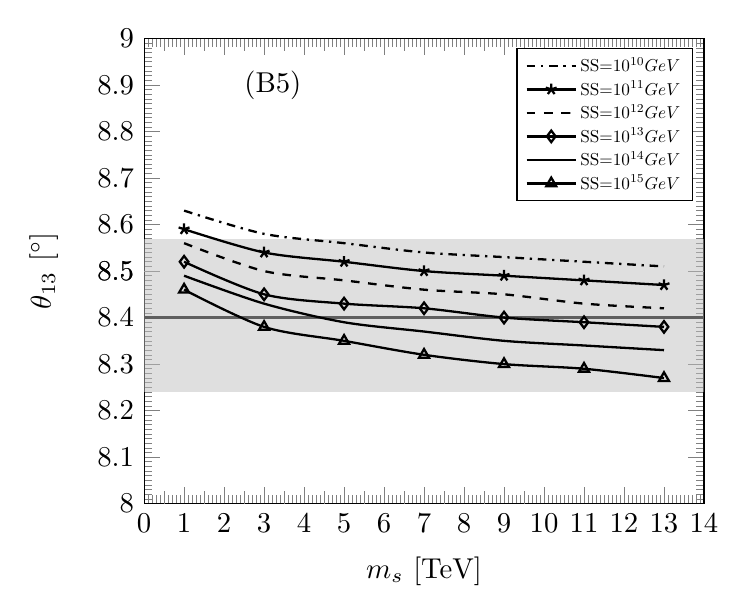}
    \subcaption{}
    \label{t135}
  \end{subfigure}
  \begin{subfigure}[b]{0.4\textwidth}
    \includegraphics[width=\textwidth]{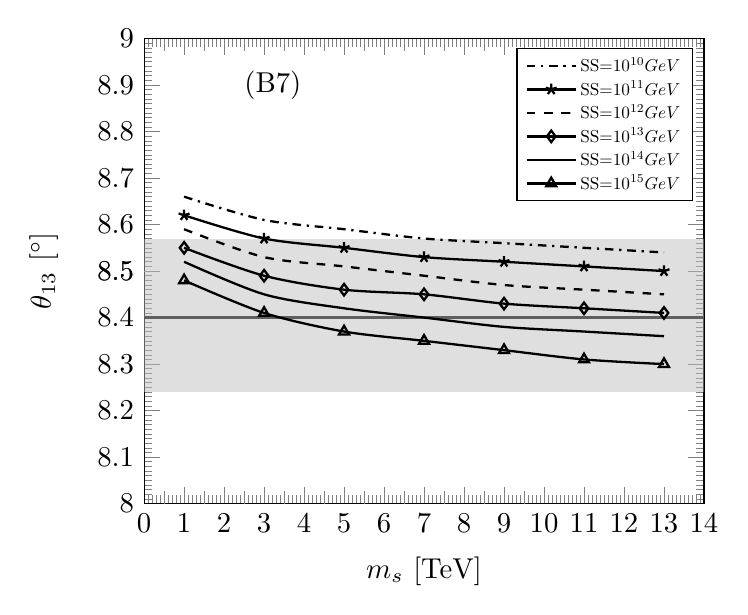}
    \subcaption{}
    \label{t137}
  \end{subfigure}
    \begin{subfigure}[b]{0.4\textwidth}
    \includegraphics[width=\textwidth]{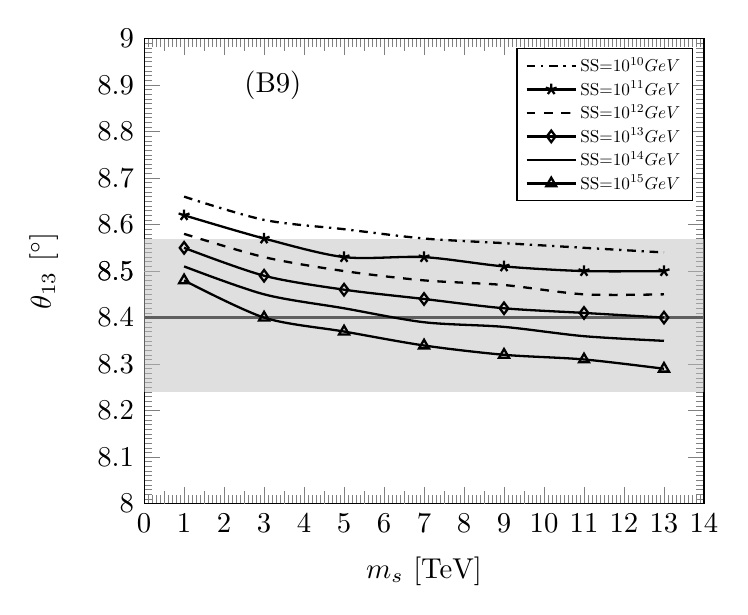}
    \subcaption{}
    \label{t139}
  \end{subfigure}
  \begin{subfigure}[b]{0.4\textwidth}
    \includegraphics[width=\textwidth]{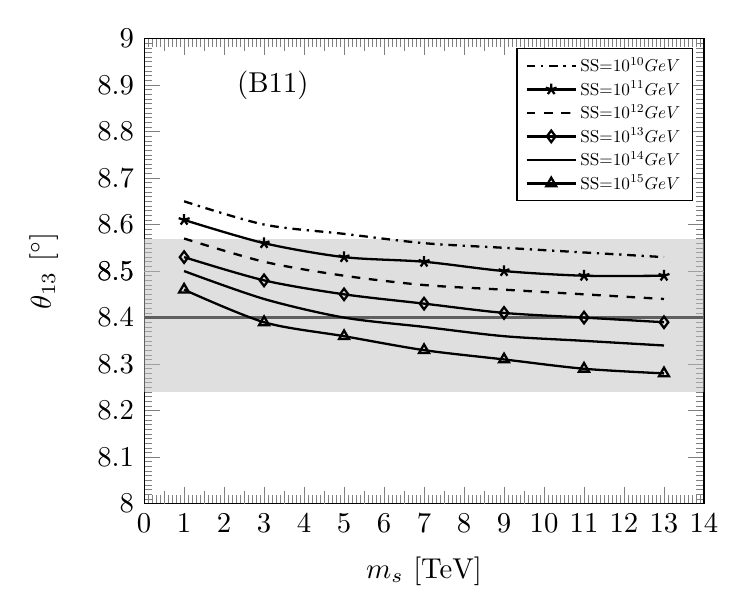}
    \subcaption{}
    \label{t1311}
  \end{subfigure}
    \begin{subfigure}[b]{0.4\textwidth}
    \includegraphics[width=\textwidth]{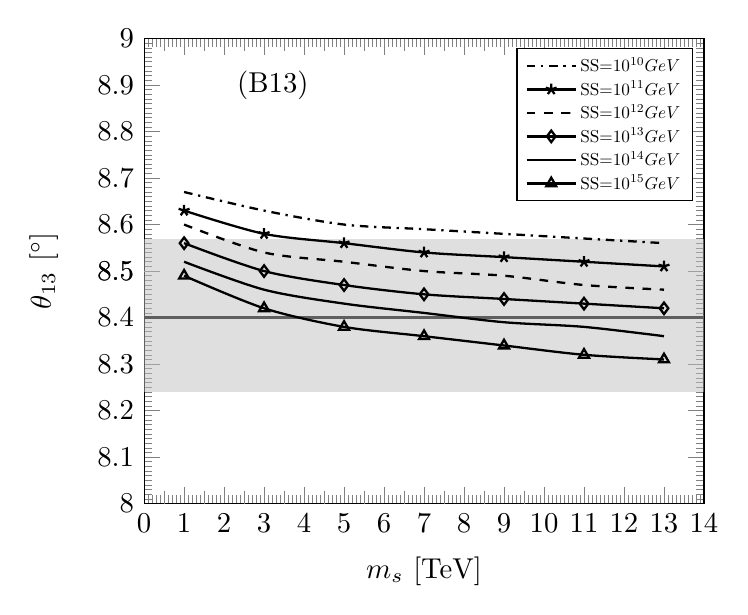}
    \subcaption{}
    \label{t1313}
  \end{subfigure}
  \caption{\scriptsize The fluctuations of the numerical values of $\theta_{13}$, at the EW scale is studied, against changing $m_s$, and the SS scale. The shaded region (horizontal) represents the experimental $3\sigma$ range \cite{deSalas:2017kay}, and the horizontal bold line inside the shaded region indicates the best-fit value. The six figures (a), (b), (c), (d), (e), and (f) are for the different input data sets B3, B5, B7, B9, B11, and B13 respectively (as given in Table.\,(\ref{90i})). The SS scales are fixed at $10^{10}$ GeV, $10^{11}$ GeV, $10^{12}$ GeV, $10^{13}$ GeV, $10^{14}$ GeV, and $10^{15}$ GeV.}
 \label{t13msss}
\end{figure}

\begin{figure}
  %
  \begin{subfigure}[b]{0.4\textwidth}
    \includegraphics[width=\textwidth]{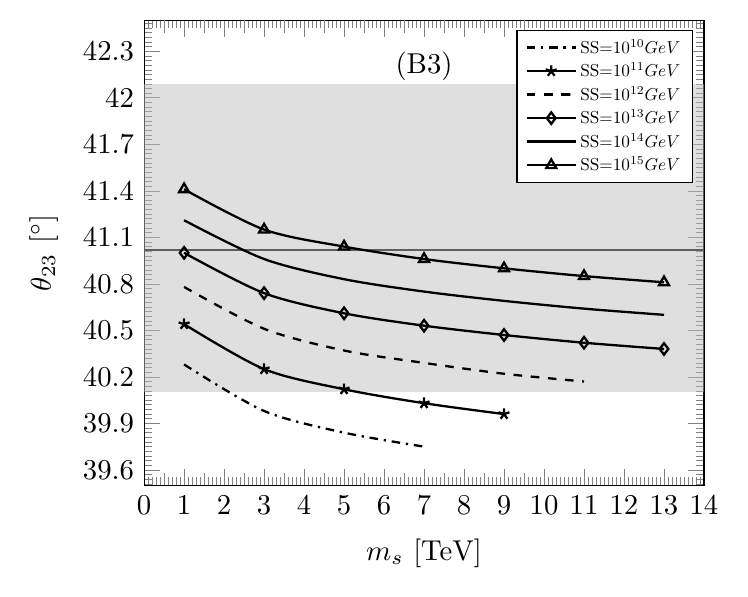}
\captionsetup[figure]{font=small,skip=0pt}
    \subcaption{}
    \label{scr3}
  \end{subfigure}
\vspace*{1mm}  
    \begin{subfigure}[b]{0.4\textwidth}
    \includegraphics[width=\textwidth]{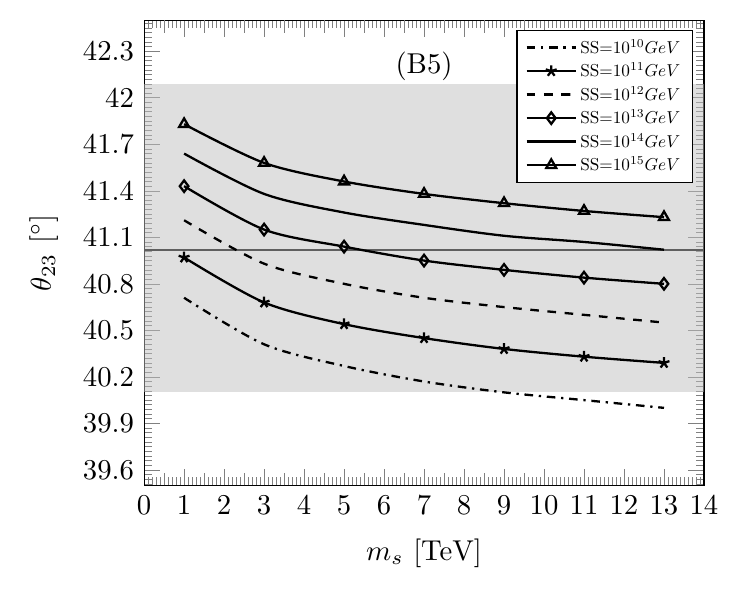}
    \subcaption{}
    \label{scr5}
  \end{subfigure}
  \begin{subfigure}[b]{0.4\textwidth}
    \includegraphics[width=\textwidth]{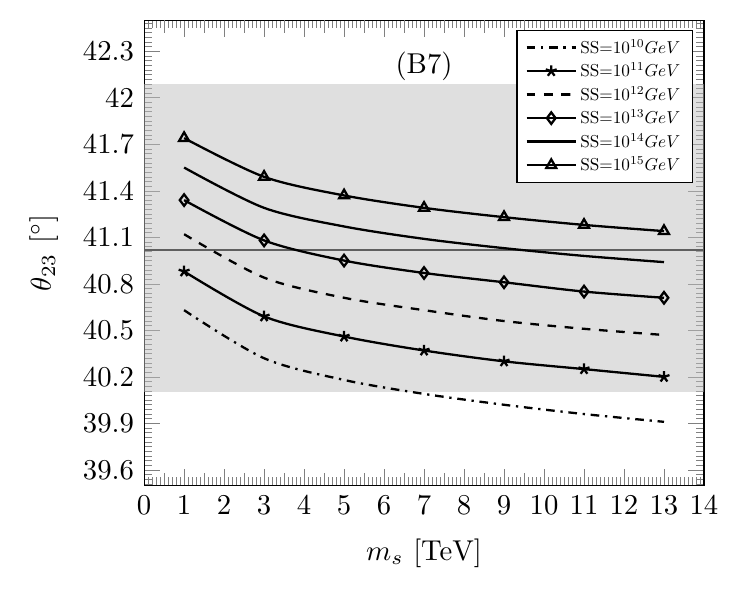}
    \subcaption{}
    \label{scr7}
  \end{subfigure}
\vspace*{1mm}  
    \begin{subfigure}[b]{0.4\textwidth}
    \includegraphics[width=\textwidth]{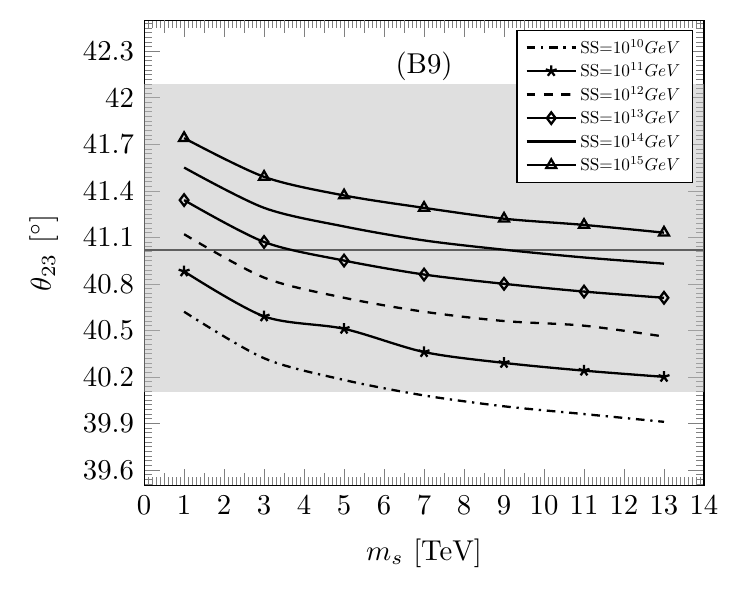}
    \subcaption{}
    \label{scr9}
  \end{subfigure}
  \begin{subfigure}[b]{0.4\textwidth}
    \includegraphics[width=\textwidth]{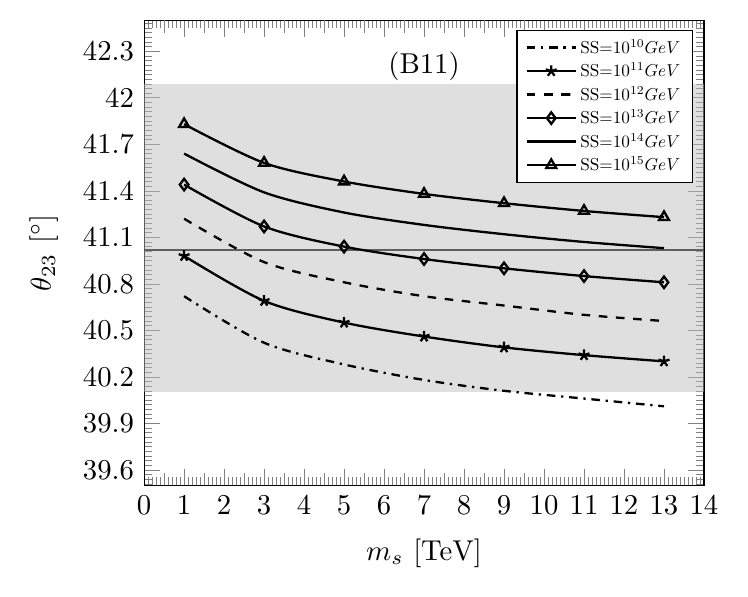}
    \subcaption{}
    \label{scr11}
  \end{subfigure}
    \begin{subfigure}[b]{0.4\textwidth}
    \includegraphics[width=\textwidth]{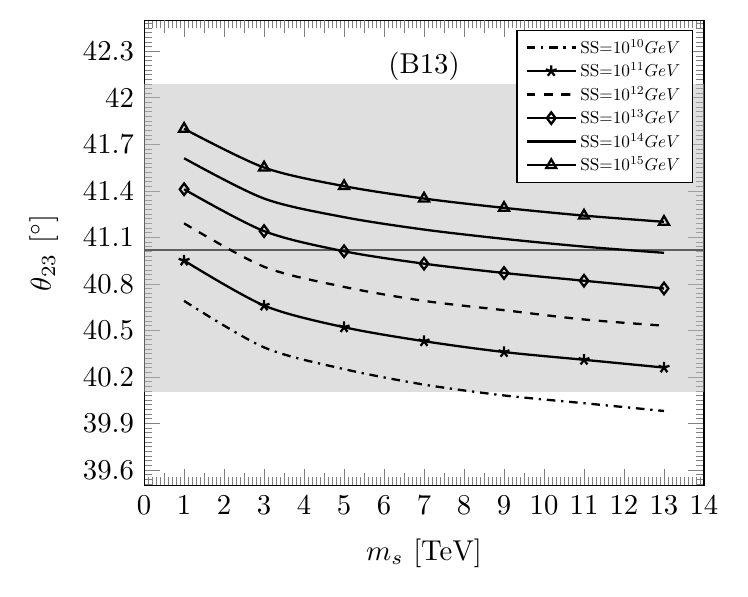}
    \subcaption{}
    \label{scr13}
  \end{subfigure}
 \caption{\scriptsize The fluctuations of the numerical values of $\theta_{23}$, at the EW scale is studied, against changing $m_s$, and the SS scale. The shaded region (horizontal) represents the experimental $3\sigma$ range \cite{deSalas:2017kay}, and the horizontal bold line inside the shaded region indicates the best-fit value. The six figures (a), (b), (c), (d), (e), and (f) are for the different input data sets B3, B5, B7, B9, B11, and B13 respectively (as given in Table.\,(\ref{90i})). The SS scales are fixed at $10^{10}$ GeV, $10^{11}$ GeV, $10^{12}$ GeV, $10^{13}$ GeV, $10^{14}$ GeV, and $10^{15}$ GeV.}
\label{t23msss}
\end{figure}

\begin{figure}
  %
  \begin{subfigure}[b]{0.4\textwidth}
    \includegraphics[width=\textwidth]{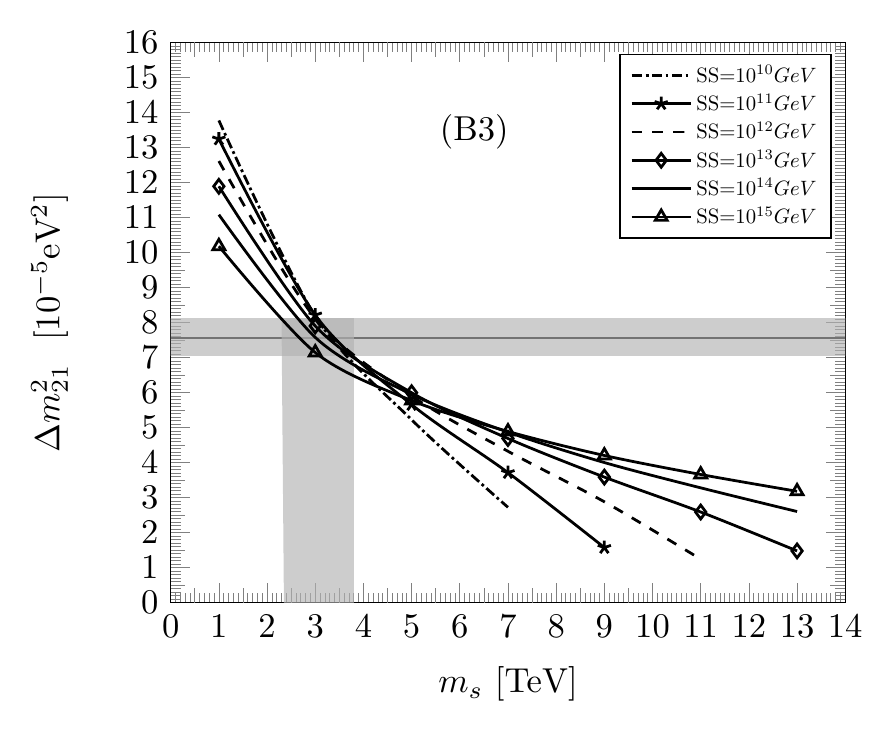}
    \subcaption{}
\label{md217msss}
  \end{subfigure}
  \begin{subfigure}[b]{0.4\textwidth}
    \includegraphics[width=\textwidth]{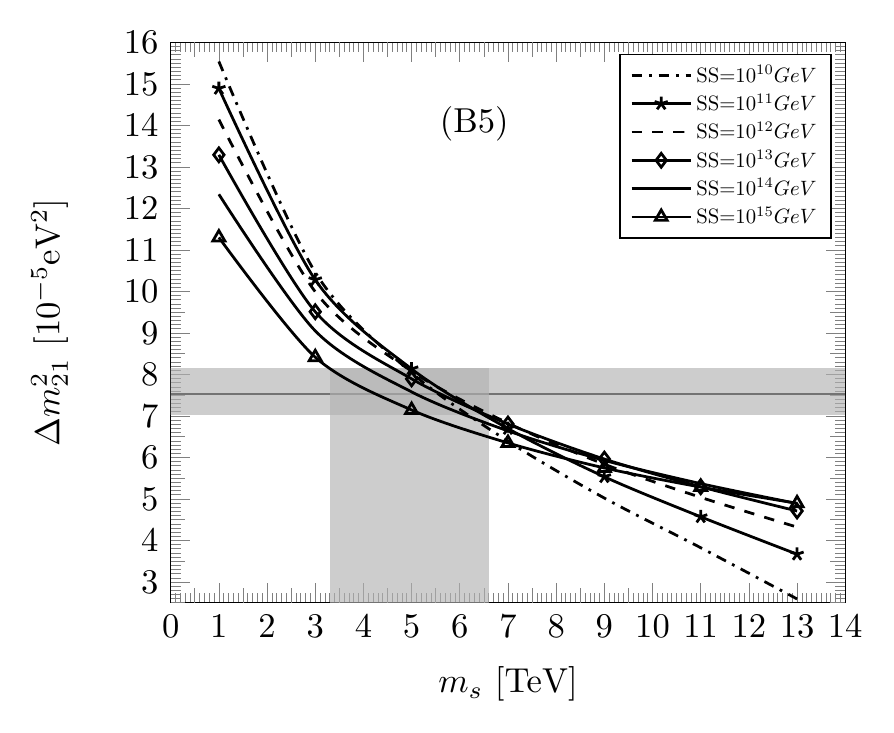}
    \subcaption{}
\label{md215msss}
  \end{subfigure}
  \begin{subfigure}[b]{0.4\textwidth}
    \includegraphics[width=\textwidth]{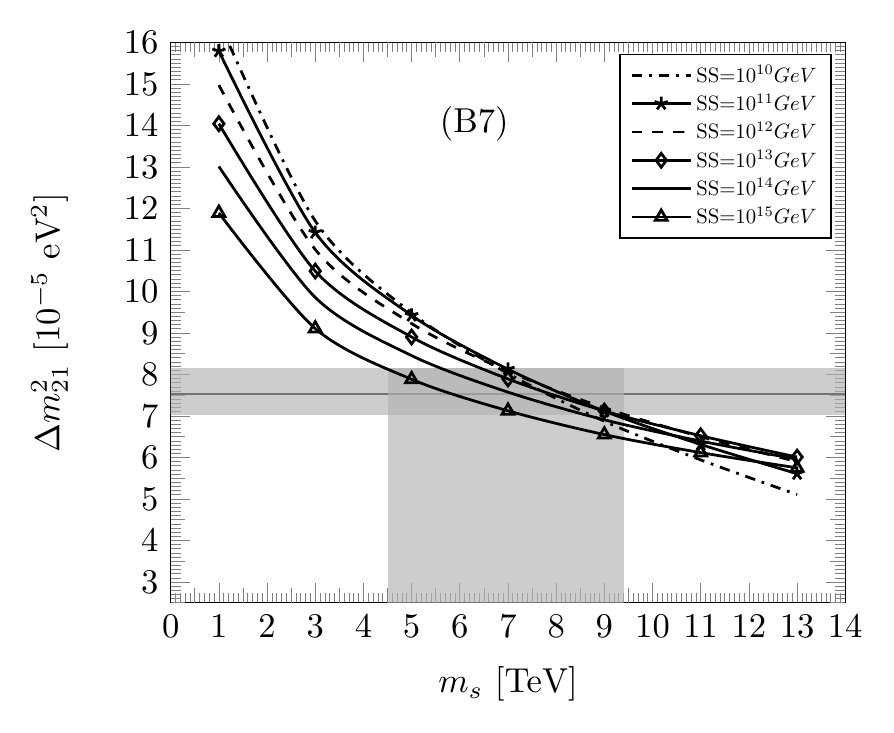}
    \subcaption{}
\label{md217msss}
  \end{subfigure}
\begin{subfigure}[b]{0.4\textwidth}
    \includegraphics[width=\textwidth]{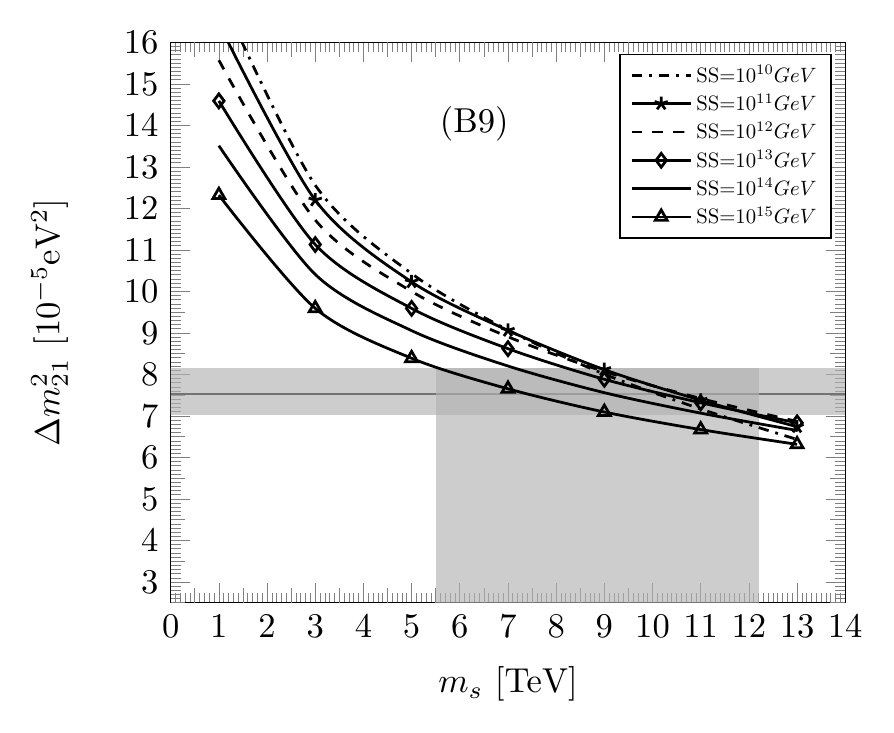}
\subcaption{}
\label{md219msss}
  \end{subfigure}
  \begin{subfigure}[b]{0.4\textwidth}
    \includegraphics[width=\textwidth]{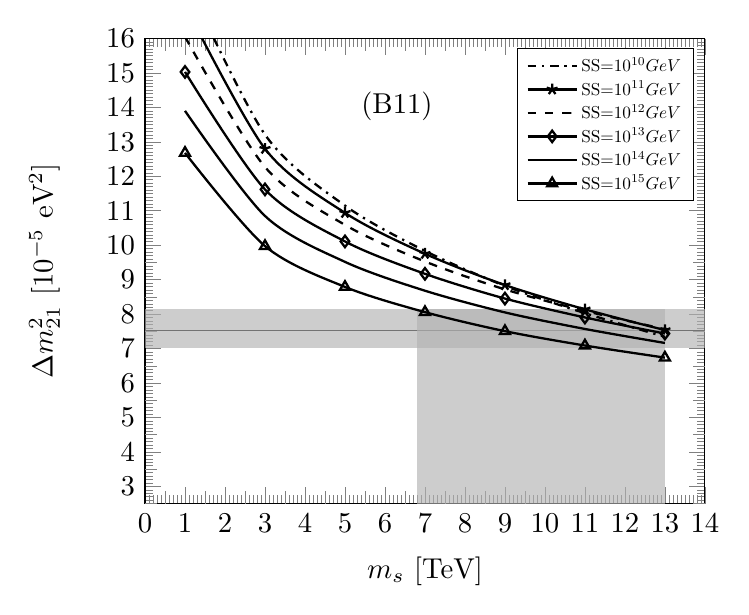}
\subcaption{}
\label{md2111msss}
  \end{subfigure}
  \begin{subfigure}[b]{0.4\textwidth}
    \includegraphics[width=\textwidth]{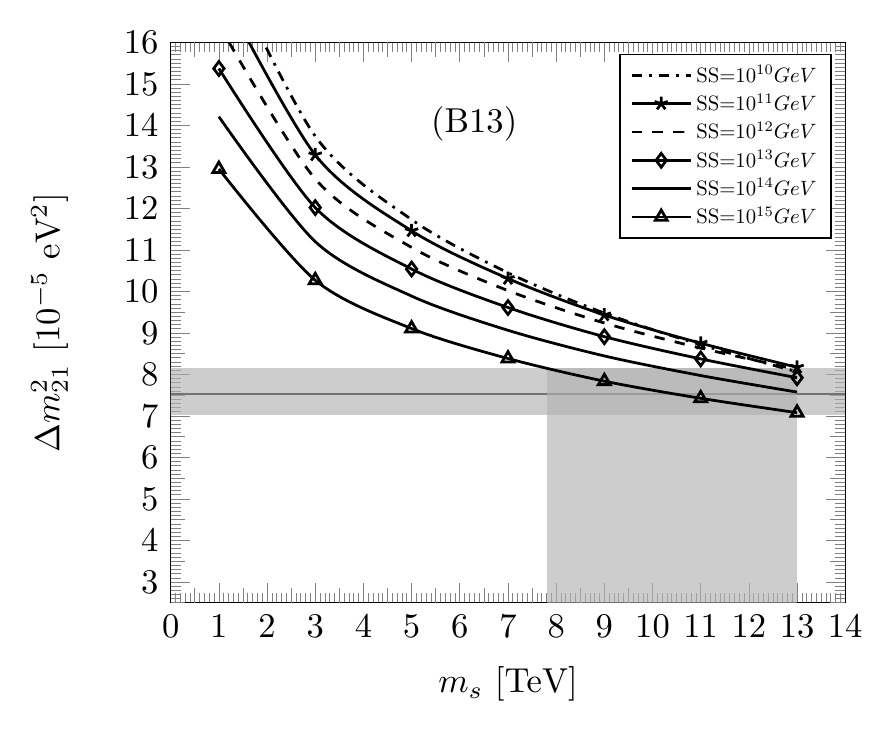}
\subcaption{}
\label{md2113msss}
  \end{subfigure}
  \captionsetup{justification=raggedright, singlelinecheck=true,  width=0.95\linewidth}
\caption{\scriptsize{ The fluctuations of the numerical values of $\Delta m_{21}^2$, at the EW scale is studied, against changing $m_s$, and the SS scale. The shaded region (horizontal) represents the experimental $3\sigma$ range \cite{deSalas:2017kay}, and the horizontal bold line inside the shaded region indicates the best-fit value. The vertical shaded region corresponds to the allowed $m_s$ region, for which the plots for different SS scale lie within the $3\sigma$ bound. The six figures (a), (b), (c), (d), (e), and (f) are for the different input data sets B3, B5, B7, B9, B11, and B13 respectively (as given in Table.\,(\ref{90i})). The SS scales are fixed at $10^{10}$ GeV, $10^{11}$ GeV, $10^{12}$ GeV, $10^{13}$ GeV, $10^{14}$ GeV, and $10^{15}$ GeV.}}
\label{md21msss}
\end{figure}

\begin{figure}
  %
  \begin{subfigure}[b]{0.4\textwidth}
    \includegraphics[width=\textwidth]{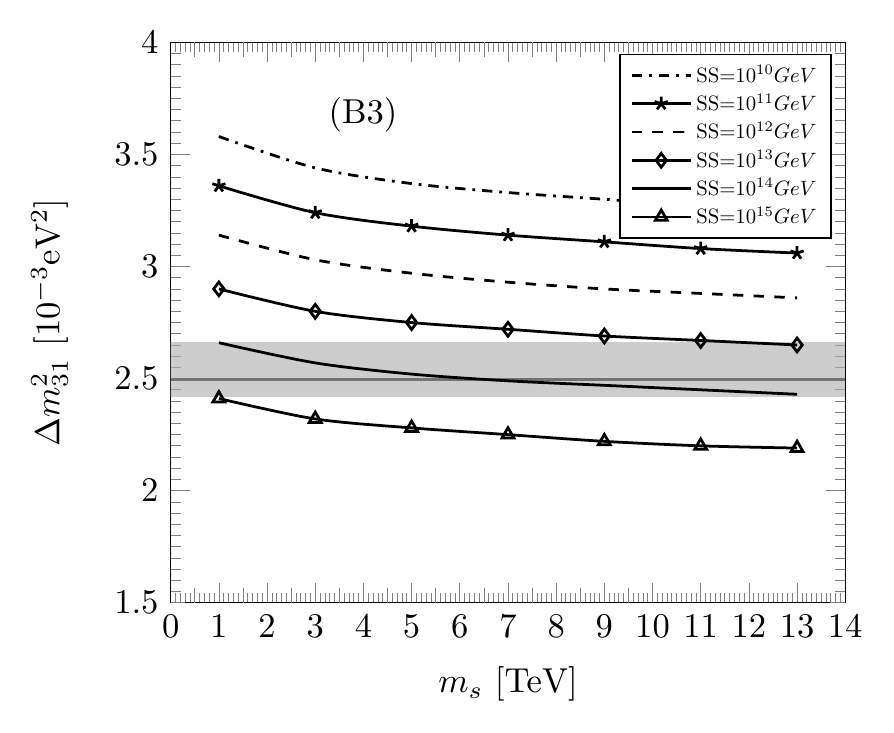}
    \subcaption{}
\label{md317msss}
  \end{subfigure}
  \begin{subfigure}[b]{0.4\textwidth}
    \includegraphics[width=\textwidth]{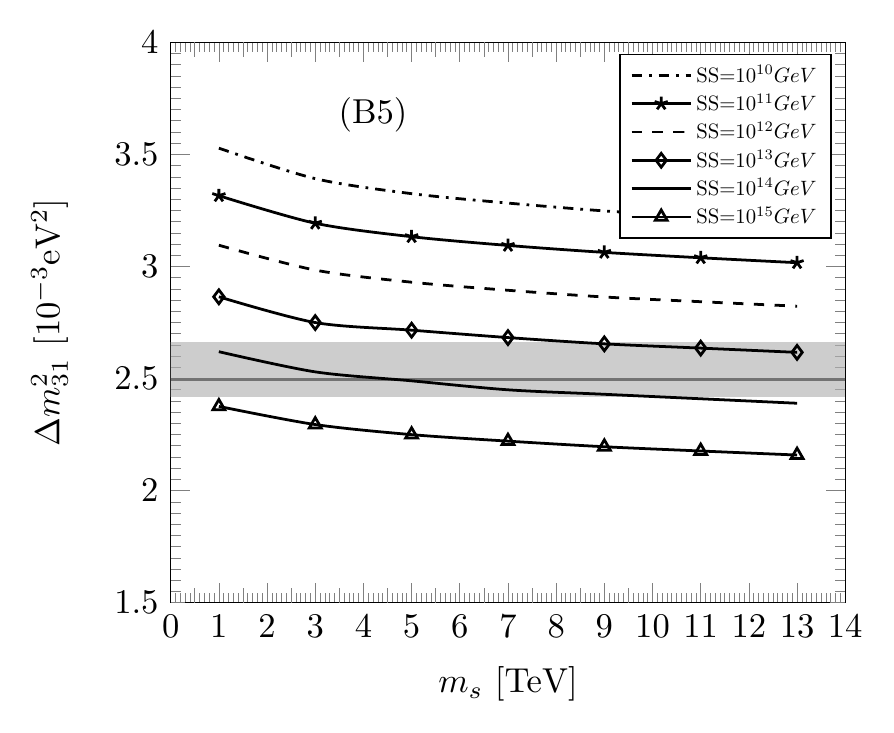}
    \subcaption{}
\label{md315msss}
  \end{subfigure}
  \begin{subfigure}[b]{0.4\textwidth}
    \includegraphics[width=\textwidth]{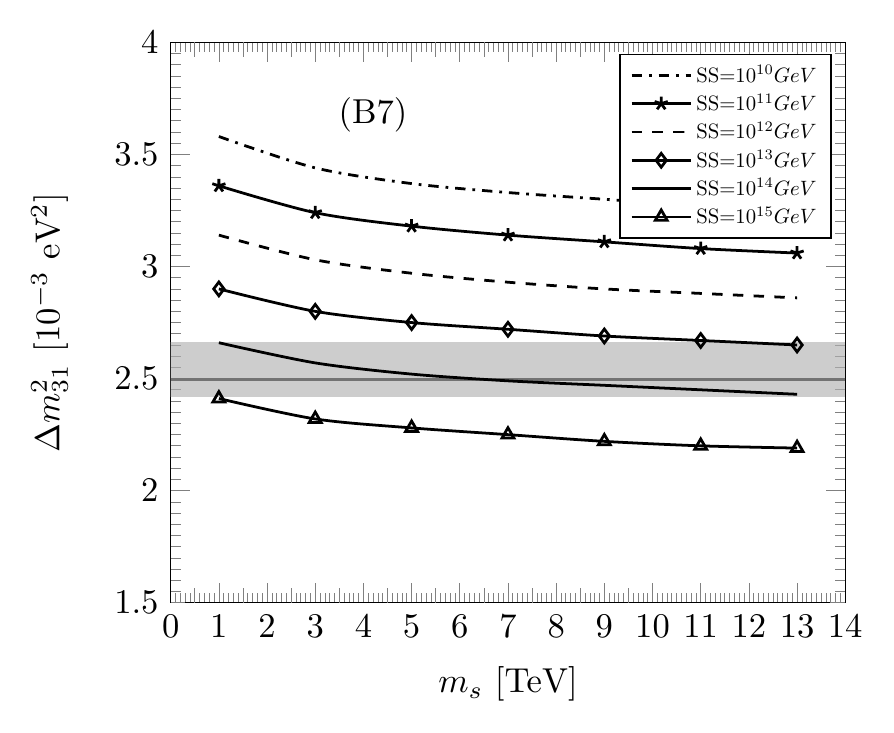}
    \subcaption{}
\label{md317msss}
  \end{subfigure}
\begin{subfigure}[b]{0.4\textwidth}
    \includegraphics[width=\textwidth]{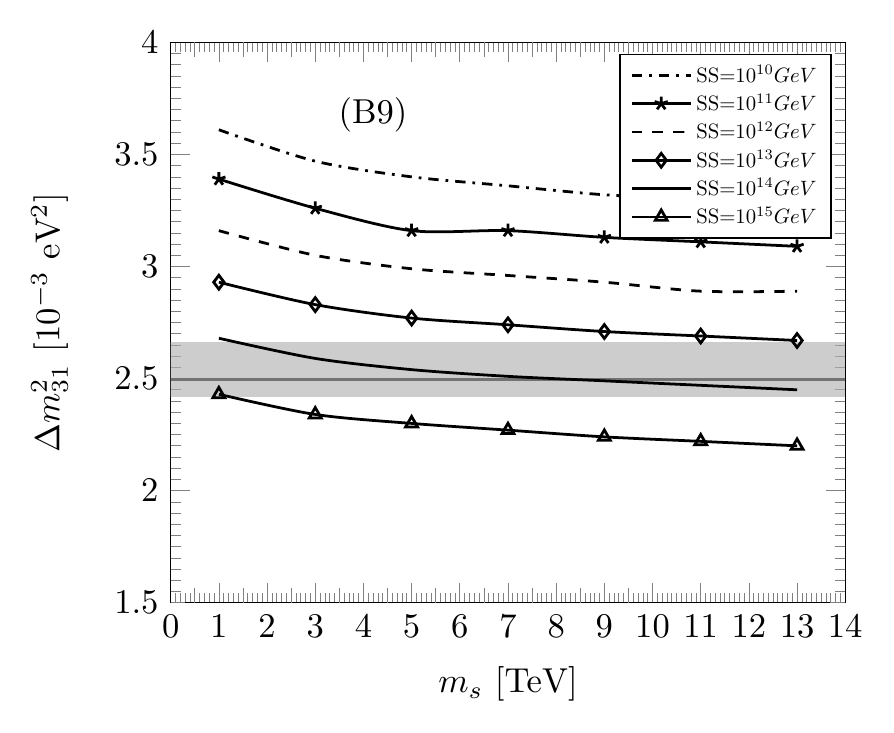}
\subcaption{}
\label{md319msss}
  \end{subfigure}
  \begin{subfigure}[b]{0.4\textwidth}
    \includegraphics[width=\textwidth]{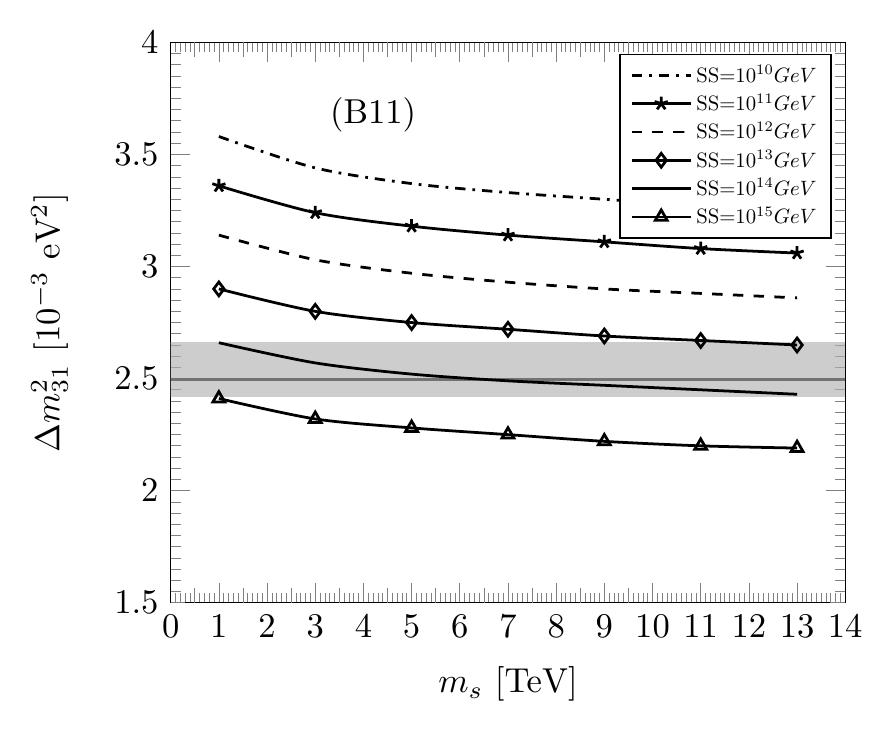}
\subcaption{}
\label{md3111msss}
  \end{subfigure}
  \begin{subfigure}[b]{0.4\textwidth}
    \includegraphics[width=\textwidth]{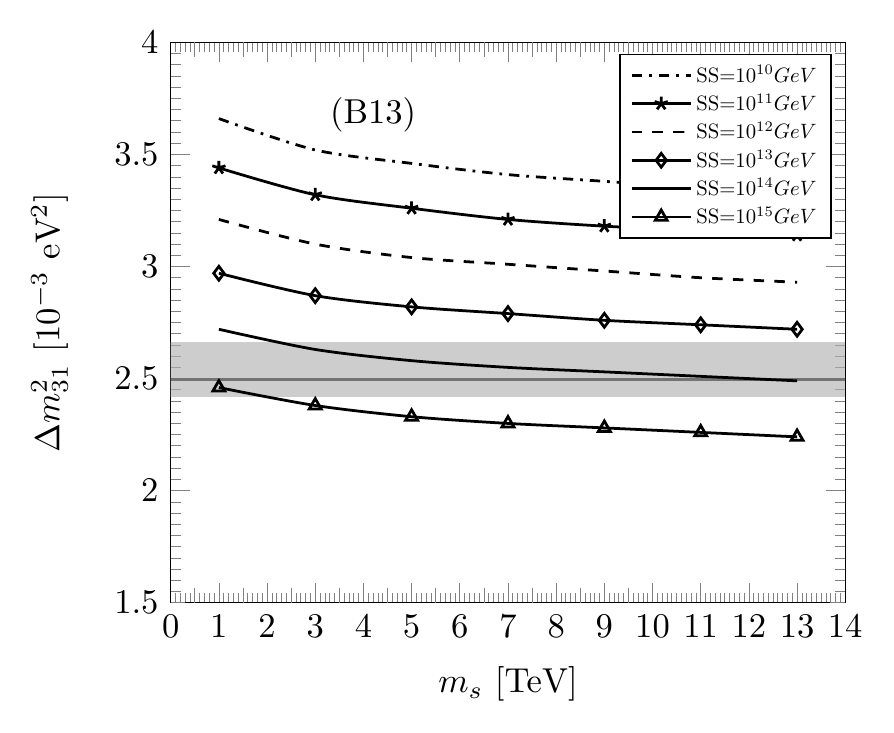}
\subcaption{}
\label{md3113msss}
  \end{subfigure}
  \captionsetup{justification=raggedright, singlelinecheck=false,  width=0.95\linewidth}
\caption{\scriptsize{ The fluctuations of the numerical values of $\Delta m_{31}^2$, at the EW scale is studied, against changing $m_s$, and the SS scale. The shaded region (horizontal) represents the experimental $3\sigma$ range \cite{deSalas:2017kay}, and the horizontal bold line inside the shaded region indicates the best-fit value. The vertical shaded region corresponds to the allowed $m_s$ region, for which the plots for different SS scale lie within the $3\sigma$ bound. The six figures (a), (b), (c), (d), (e), and (f) are for the different input data sets B3, B5, B7, B9, B11, and B13 respectively (as given in Table.\,(\ref{90i})). The SS scales are fixed at $10^{10}$ GeV, $10^{11}$ GeV, $10^{12}$ GeV, $10^{13}$ GeV, $10^{14}$ GeV, and $10^{15}$ GeV.}}
\label{md31msss}
\end{figure}

\begin{enumerate}
\item[(a)] Among the three mixing angles, $\theta_{13}$ at EW scale decreases if the SS scale is increased, whereas $\theta_{12}$ and $\theta_{23}$ increase. For wide ranges of $m_s$ and the SS scale, the output values stay within the $3\sigma$ bound. However, for different input data sets concerned, the exclusion of certain $m_{s}$ values or SS scales are also possible, depending upon the $3\sigma$ bound of the concerned mixing angles. For example, consider the case of $\theta_{12}$ at the EW scale, against a fixed input data set $B5$. If we believe SS scale to be $10^{10}$\,GeV, then, from the plots, it is evident that the  susy breaking scale should not be more than 7\,TeV (See Figs.(\ref{t12msss})).  For the other two mixing angles, ($\theta_{13}$), and ($\theta_{13}$) see Figs.\ref{t13msss} and \ref{t23msss} respectively.

\item[(b)] With all the conditions being the same as before, the $\delta$ increases if the SS scale is increased, whereas the reverse is true for the Majorana phases. (See Figs.\,(\ref{dmsss})-(\ref{pmsss})).

\item[(c)] We observe certain interesting results in concern with $\Delta m_{21}^2$ and $\Delta m_{31}^2$. The mass squared differences are found highly  sensitive to the initial data set, $m_s$ and the SS scale. The $\Delta m_{31}^2$ remains more or less stable against $m_{s}$, but crosses $3\sigma$ bound if SS scale is varied. On the contrary, the $\Delta m_{21}^2$ fluctuates more with $m_s$, but less with SS scale. It is interesting to note that against a fixed input data set (say, B5), with respect to $3\sigma$ range of $\Delta m_{21}^2$, one can even find a bound over the $m_{s}$ scale. This bound shifts to the right, i.e, towards a higher $m_s$ region as we take the input numerals as per the initial data sets from $B1$ to $B13$ (see Figs. \ref{md21msss} and \ref{md31msss}).    
\end{enumerate}

\subsection{The SC relation and the mass ratios}
In addition to the physical observables, we try to see how the certain parameters/ relation  evolve against the varying energy scale. The neutrino oscillation experiments hint not for individual neutrino masses, the study of individual parameters and how they evolve carry physical insight. This study is relevant from the model building point of view.

\begin{figure}
  %
  \begin{subfigure}[b]{0.4\textwidth}
    \includegraphics[width=\textwidth]{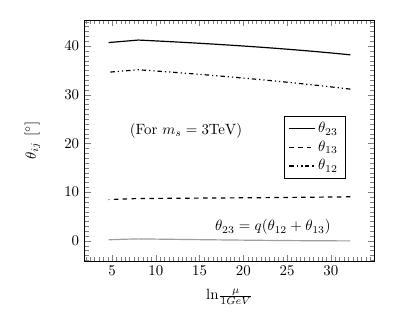}
    \subcaption{}
    \label{scr3}
  \end{subfigure}
\vspace*{1mm}  
    \begin{subfigure}[b]{0.4\textwidth}
    \includegraphics[width=\textwidth]{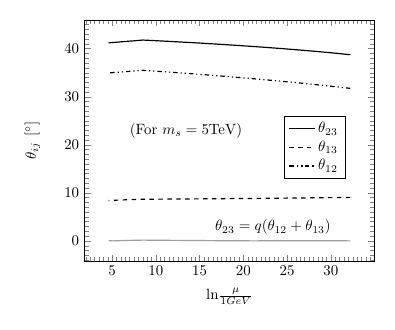}
    \subcaption{}
    \label{scr5}
  \end{subfigure}
  \begin{subfigure}[b]{0.4\textwidth}
    \includegraphics[width=\textwidth]{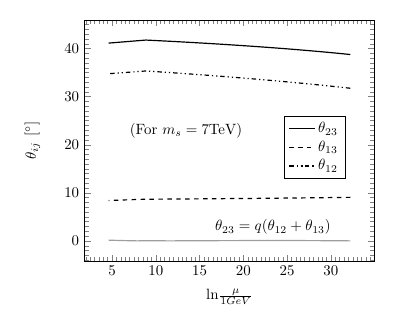}
    \subcaption{}
    \label{scr7}
  \end{subfigure}
\vspace*{1mm}  
    \begin{subfigure}[b]{0.4\textwidth}
    \includegraphics[width=\textwidth]{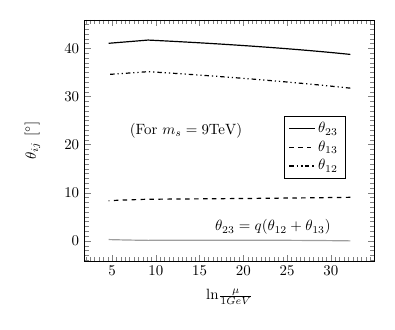}
    \subcaption{}
    \label{scr9}
  \end{subfigure}
  \begin{subfigure}[b]{0.4\textwidth}
    \includegraphics[width=\textwidth]{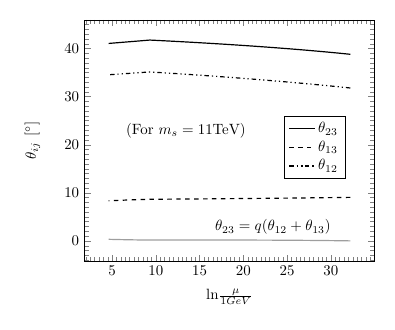}
    \subcaption{}
    \label{scr11}
  \end{subfigure}
    \begin{subfigure}[b]{0.4\textwidth}
    \includegraphics[width=\textwidth]{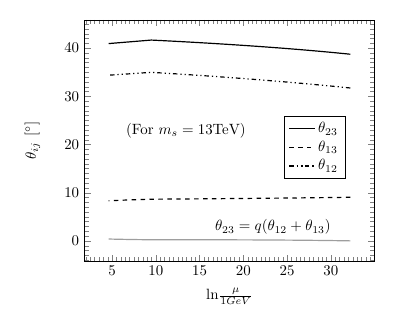}
    \subcaption{}
    \label{scr13}
  \end{subfigure}
  \caption{\scriptsize Radiative evolution of the three neutrino mixing angles and its self-complementarity relation from the seesaw scale to the EW scale for different choices of $m_s$ are studied. The six figures (a), (b), (c), (d), (e), and (f) are for the different input data sets B3, B5, B7, B9, B11, and B13 respectively (as given in Table.\,(\ref{90i})). Here we consider only one SS scale ($10^{14}$ GeV).}
  \label{scrfbs}
\end{figure}

\begin{figure}
  \begin{subfigure}[b]{0.4\textwidth}
    \includegraphics[width=\textwidth]{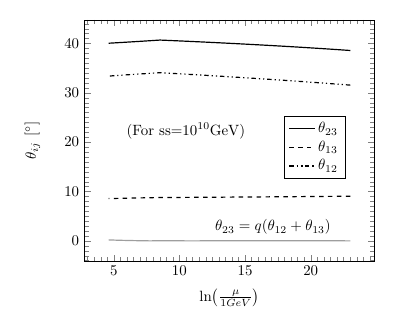}
    \subcaption{}
    \label{scrss10}
  \end{subfigure}
  \begin{subfigure}[b]{0.4\textwidth}
    \includegraphics[width=\textwidth]{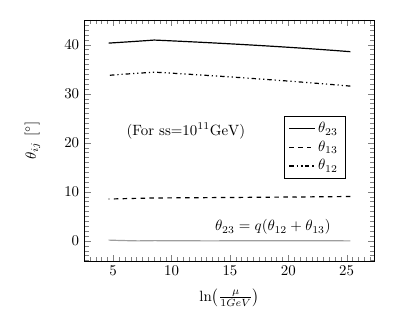}
    \subcaption{}
    \label{scrss11}
  \end{subfigure}
  
    \begin{subfigure}[b]{0.4\textwidth}
    \includegraphics[width=\textwidth]{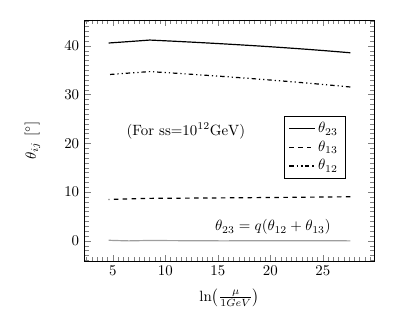}
    \subcaption{}
    \label{scrss12}
  \end{subfigure}
  \begin{subfigure}[b]{0.4\textwidth}
    \includegraphics[width=\textwidth]{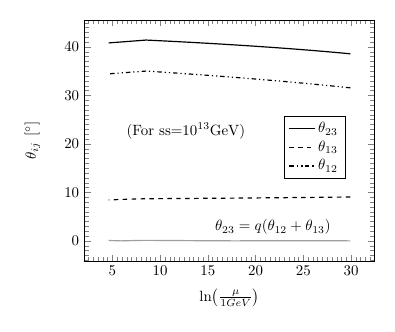}
    \subcaption{}
    \label{scrss13}
  \end{subfigure}
  
    \begin{subfigure}[b]{0.4\textwidth}
    \includegraphics[width=\textwidth]{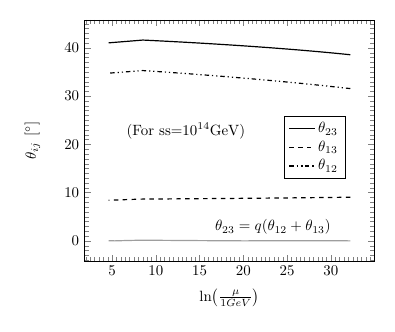}
    \subcaption{}
    \label{scrss14}
  \end{subfigure}
  \begin{subfigure}[b]{0.4\textwidth}
    \includegraphics[width=\textwidth]{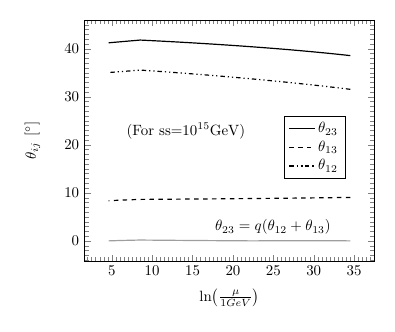}
    \subcaption{}
    \label{scrss15}
  \end{subfigure}
  \captionsetup{justification=raggedright, singlelinecheck=false,  width=0.95\linewidth}
  \caption{\scriptsize Radiative evolution of the three neutrino mixing angles and its self-complementarity relation from the seesaw scale to the EW scale for a fixed data set B5, $m_s=5$ TeV (as given in Table.\,(\ref{90i})) are studied for different seesaw scales. The six figures (a), (b), (c), (d), (e), and (f) correspond to the different choices of SS at $10^{10}$ GeV, $10^{11}$ GeV, $10^{12}$ GeV, $10^{13}$ GeV, $10^{14}$ GeV, and $10^{15}$ GeV respectively.}
  \label{scrss}
\end{figure}

\begin{figure}
  \begin{subfigure}[b]{0.4\textwidth}
    \includegraphics[width=\textwidth]{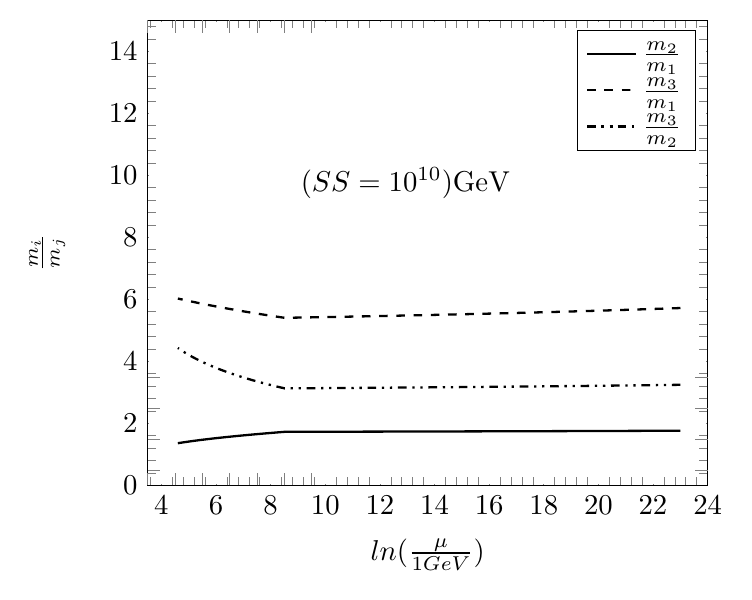}
    \subcaption{}
    \label{mr10}
  \end{subfigure}
  \begin{subfigure}[b]{0.4\textwidth}
    \includegraphics[width=\textwidth]{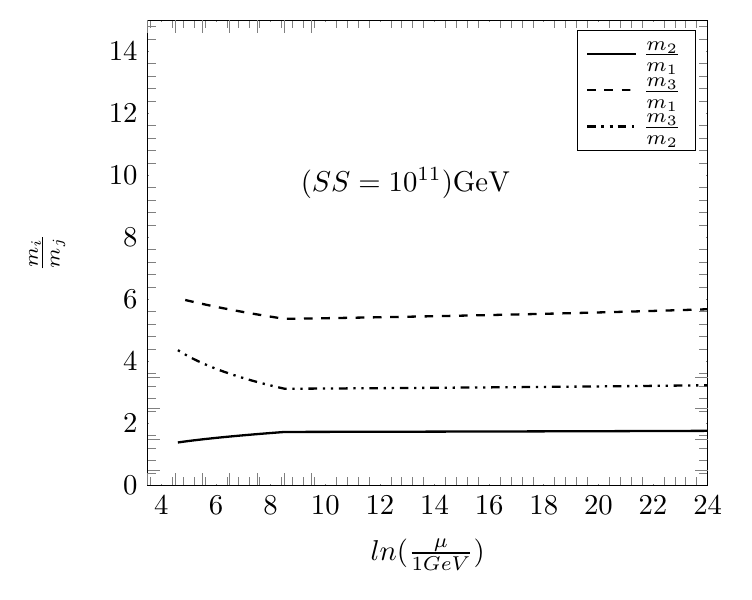}
    \subcaption{}
    \label{mr11}
  \end{subfigure}
  
    \begin{subfigure}[b]{0.4\textwidth}
    \includegraphics[width=\textwidth]{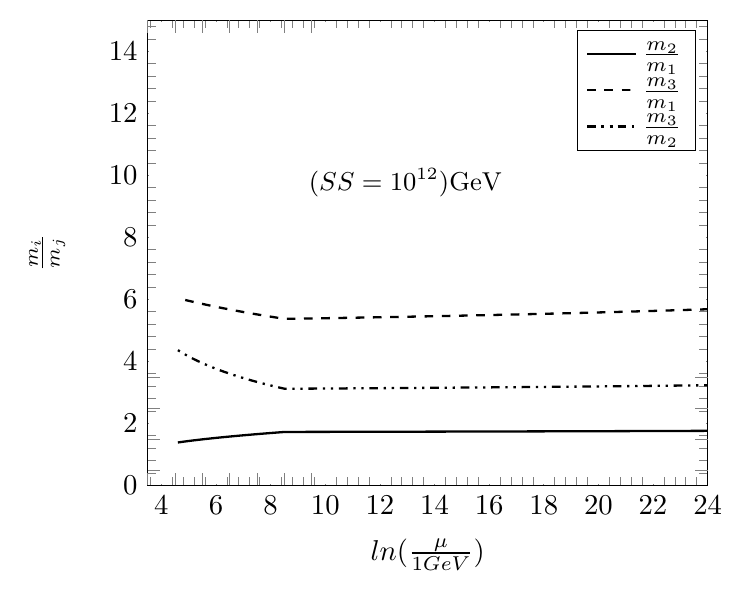}
    \subcaption{}
    \label{mr12}
  \end{subfigure}
  \begin{subfigure}[b]{0.4\textwidth}
    \includegraphics[width=\textwidth]{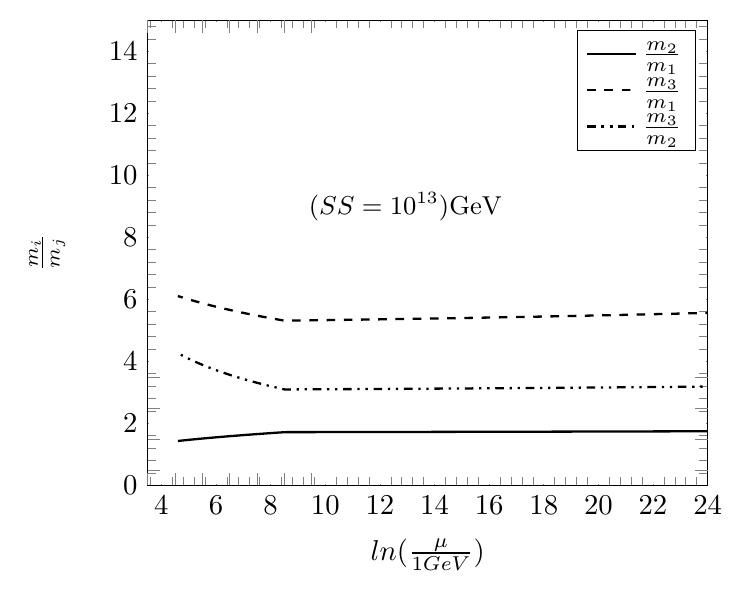}
    \subcaption{}
    \label{mr13}
  \end{subfigure}
  
    \begin{subfigure}[b]{0.4\textwidth}
    \includegraphics[width=\textwidth]{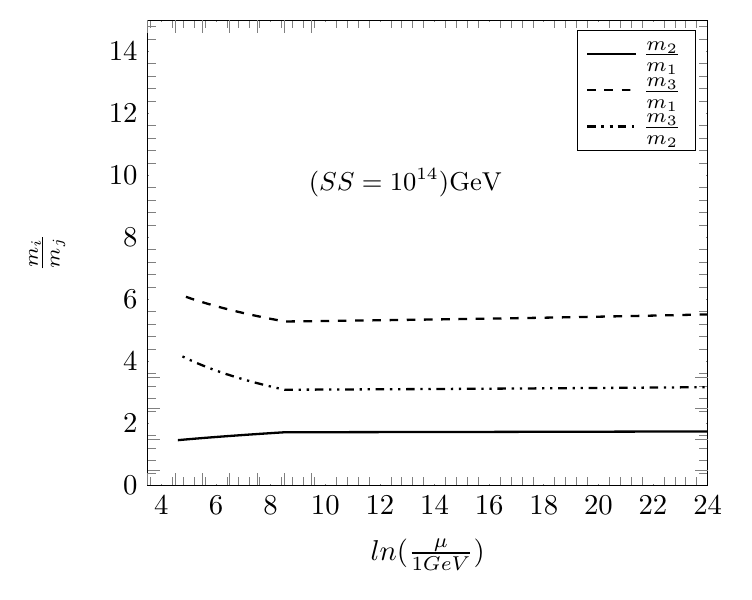}
    \subcaption{}
    \label{mr14}
  \end{subfigure}
  \begin{subfigure}[b]{0.4\textwidth}
    \includegraphics[width=\textwidth]{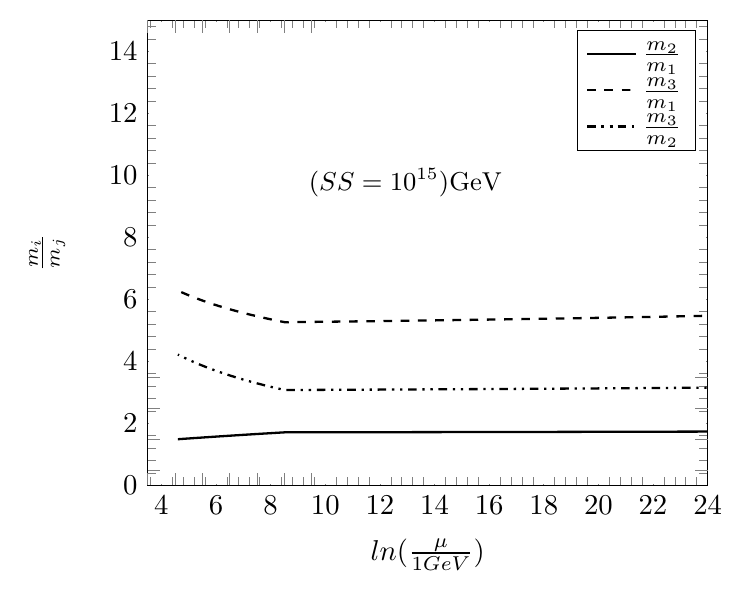}
    \subcaption{}
    \label{mr15}
  \end{subfigure}
  \captionsetup{justification=raggedright, singlelinecheck=false,  width=0.95\linewidth}
  \caption{\scriptsize Radiative evolution of the three neutrino mass ratios from the seesaw scale to the EW scale for a fix input data set B5, fix $m_s=5$ TeV (as given in Table.\,(\ref{90i})) for different seesaw scales are studied. The six figures (a), (b), (c), (d), (e), and (f) correspond to the different choices of SS at $10^{10}$ GeV, $10^{11}$ GeV, $10^{12}$ GeV, $10^{13}$ GeV, $10^{14}$ GeV, and $10^{15}$ GeV respectively.}
  \label{mr}
\end{figure}

\begin{enumerate}
\item[(a)]As stated earlier, we have assumed that at the SS scale ,the three mixing angles are connected via a complementarity relation (See Eq.\,(\ref{comp2})). We see that for a fixed $m_s$ and a chosen SS scale, with all the input parameters fixed to a certain data set (say,B5), the angles evolve (except $\theta_{13}$ which is almost stable), but the SC relation connecting the mixing angles, remains almost invariant against the radiative evolution. This stability is achievable, even if we vary the SS scale or $m_{s}$. We have shown the radiative evolution of the angles along with the SC relation for both varying $m_{s}$ (with a fixed SS scale) and varying SS scale (with fixed $m_s$). For details, see Figs.\,(\ref{scrfbs})-(\ref{scrss}). The SC relation is a phenomenologically motivated relation like the QLC relation\cite{Zhang:2012pv} that connects the quark and lepton sectors. A relation of this kind bears the signature of a certain hidden symmetry. As pointed out in our analysis, that which reflects the invariance of the former against radiative evolution may turn out as a fruitful information for the model builders.  

\item[(b)] Like the mixing angles, we try to see how the mass parameters respond to radiative evolution. Instead of concentrating on individual neutrino masses, we focus on the three mass ratios as such: $m_{2}/m_1$, $m_{3}/m_1$, and $m_{3}/m_2$. This is inspired by the phenomenology of the quark sector. Where, we see that the mass ratio between the down and strange quarks is naturally related to the quark mixing angle (Cabibbo angle) which plays an important role in describing the mixing among the quarks\cite{Araki:2013rkf,Wang:2014sba}. To exemplify, we fix the $m_{s}$ at $5$\,TeV and the input data set at $B5$. Following this, we see how the three neutrino mass ratios vary against the changing SS scale. The details are shown in Fig.(\ref{mr}). One sees that the ratio $m_{3}/m_{1}$ or $m_{3}/m_2$, though remains invariant in the SUSY region, yet changes after crossing the $m_s$ scale. But, interestingly, the ratio $m_{2}/m_{1}$ remains almost invariant and tries to maintain a constant numerical value as such: $m_{2}/m_1\sim 2$. A summarized version of the different types of effect each neutrino parameters receive due to the variation of $m_s$ and SS are given in Table \ref{eff}.

\end{enumerate}

\section{Summary and Discussion}

In this paper we have studied the radiative evolution of neutrino observational  parameters for varying $m_s$ scale following a top-down approach. We presume the hierarchy of the three neutrino masses to be of normal type. All the nine observational parameters related to neutrino oscillations are allowed to run down from the seesaw scale up to the electroweak scale using their respective RGEs (both MSSM and SM). We also use the RGEs of the three gauge couplings, third generation Yukawa couplings, and quartic Higgs coupling. All the neutrino parameters along with  the other couplings undergo RG evolution, and subsequently, get different RG corrections. The $m_s$, which appears to be a leading parameter is kept varying between 1\,TeV to 13\,TeV and the effect of such a variation on the observational parameters at the EW scale is noted. Instead of adhering to a fixed SS scale, we allow the latter to change between $10^{10}$\,GeV to $10^{15}$\,GeV, and have checked how the observational parameters vary. Besides, the work reveals that the self-complementarity relation among the mixing angles remains stable against the radiative evolution. Also, we have studied how certain parameters like neutrino mass ratios behave during this evolution.  

The relevance of the SUSY is unavoidable in the context of particle physics, as it can answer to certain important theoretical issues like the hierarchy problem, the unification of gauge couplings, the existence of dark matter etc. But, unlike the Standard Model, the SUSY is still lacking the experimental evidences. Although the LHC experiment is running at 13 TeV, it has not yet witnessed any signature of SUSY. This may imply that the SUSY breaks at certain higher energy scale which is not yet achieved by the LHC experiment, or even if it breaks at a low energy, the beam luminosity available in the LHC experiment is not sufficient to detect the same. Hence, there is still a hope that SUSY exists. The SUSY breaking scale, $m_s$, is an important parameter and influences the neutrino observational parameters. The origin of a neutrino mass owes to the seesaw mechanism, and the scale at which the latter occurred is also unknown. But theoretically one may predict that scale to be lying within the range of $10^{10}$ to $10^{15}$\,GeV. In our analysis, these two parameters, the $m_s$ and the SS scale partake a lot. Besides, the input data set (like, $Aj$ or $Bj$) which are although model independent, plays an important role. Initially, the input parameters in the data sets are chosen such that against a fixed $m_s$ and a fixed SS scale ($10^{14}$\,GeV), the neutrino observational parameters at the EW scale lie within the $3\sigma$ bound. This is to be mentioned that the initial entries in terms of the three mixing angles, follow a self-complementarity relation.

At EW scale, the three mixing angles, CP violating phases, and $\Delta m_{31}^2$  try to maintain more or less stability with respect to the $3\sigma$ bound if the $m_s$ scale is varied at a fixed SS scale. But the parameter $\Delta m_{21}^2$ is less stable at lower $m_{s}$, whereas the stability increases towards higher $m_s$. Similar stability is achievable for the three mixing angles if the SS scale is varied. But for $\Delta m_{31}^2$, the stability is lost. One sees that if the stability of $\Delta m_{21}^2$ is obtained towards a higher $m_{s}$, ruling out of a certain SS scale is possible in the light of $3\sigma$ bound of $\Delta m_{21}^2$. It is worth mentioning that a strong conclusion in view of the optimization of the SUSY breaking and SS scales can not be drawn by observing the plots (See Figs.(\ref{md21msss})-(\ref{md31msss})), because the $\Delta m_{21}^2$ at the EW scale is very much sensitive to the initial arbitrary model independent entries available in the data sets ($Aj$ and $Bj$). Justifying these initial entries under a certain model or framework goes beyond the scope of this article. But through our analysis, one can at least visualize the interplay between the $m_s$ and the SS scale and how these affects the final physical observables. Though in the present analysis, we limit ourselves not to invoke the model dependent ground of these data sets, yet we emphasize on the certain traits that these numerals may carry. We see that the data set are characterized by the SC relation, $\theta_{13}+\theta_{12}\approx\theta_{23}$ and a mass ratio: $m_{2}/m_1\sim 2$, which remains almost invariant against radiative evolution. Besides, we have observed the other mass ratios like $m_{3}/m_{1}$ or $m_{3}/m_2$, also triy to maintain a stability up to SUSY breaking scale, but after that they change. This study is motivated in the context of the quark sector, where the quark mass ratio $m_{down}/m_{strange}$ plays an important role in describing the quark mixing. Relations among certain parameters and their stability during radiative evolution may bear the traits of a certain hidden symmetry present in the lepton sector and may serve as a key to some new models.

The present study is devoted to a simple visualization, concerning the interplay between the $m_s$ and SS scale and its effect on the physical observables and certain phenomenological relations. The two  Majorana phases are not yet been measured experimentally, and to simplify the analysis, we have considered both of them as equal. Again, we have restricted ourselves only to the normal hierarchy of neutrino masses. The consideration of a  degenerate spectrum for all sparticles that we have adopted in our work, is an idealized situation and is true if $m_{s}\gg m_{t},m_{Z}$\,\cite{Arnowitt:1992aq,Yamada:1992kv}. In principle, a general study can be made by minimizing the number of assumptions in order to get a more generalized result.

\section*{Appendix}
\subsection*{RGEs for gauge couplings}
The two loop renormalization group equations for gauge couplings are \cite{Barger:1992ac,Deshpande:1994qm,Singh:1998dh,Parida:1997wx} as follows:
\begin{equation}
\frac{dg_i}{dt}=\frac{b_i}{16\pi^2}g_i^3+\left(\frac{1}{16\pi^2}\right)^2\left[\sum_{j=1}^3b_{ij} \ g_i^3 \ g_j^2 - \sum_{j=t, b,  \tau}a_{ij} \ g_i^3 \ h_j^2\right], \label{grge}
\end{equation}

where, $t=ln \mu$ and  $b_i, \ b_{ij}, \ a_{ij}$ are $\beta$ function coefficients in MSSM,
\[
	\begin{matrix}
	b_i 
	\end{matrix}
	\begin{matrix}
	=
	\end{matrix}
	\begin{pmatrix}
	6.6, 1.0, -3.0 
	\end{pmatrix}
	\begin{matrix}
	, \hspace{0.3cm} b_{ij} =
	\end{matrix}
	\begin{pmatrix}
	7.96 & 5.40 & 17.60 \\
	1.80 & 25.00 & 24.00 \\
	2.20 & 9.00 & 14.00  
	\end{pmatrix}
	\begin{matrix}
	, \hspace{0.3cm} a_{ij} =
	\end{matrix}
	\begin{pmatrix}
	5.2 & 2.8 & 3.6 \\
	6.0 & 6.0 & 2.0 \\
	4.0 & 4.0 & 0.0  
	\end{pmatrix}
	\] 
	and, for non-supersymmetric case, we have
\[
	\begin{matrix}
	b_i 
	\end{matrix}
	\begin{matrix}
	=
	\end{matrix}
	\begin{pmatrix}
	4.100, -3.167, -7.000
	\end{pmatrix}
	\begin{matrix}
	, \hspace{0.3cm} g_{ij} =
	\end{matrix}
	\begin{pmatrix}
	3.98 & 2.70 & 8.8 \\
	0.90 & 5.83 & 12.0 \\
	1.10 & 4.50 & -26.0  
	\end{pmatrix}
	\begin{matrix}
	, \hspace{0.3cm} a_{ij} =
	\end{matrix}
	\begin{pmatrix}
	0.85 & 0.5 & 0.5 \\
	1.50 & 1.5 & 0.5 \\
	2.00 & 2.0 & 0.0  
	\end{pmatrix}.
	\]	

\subsection*{RGEs for Yukawa couplings}
At two-loop level for MSSM, \cite{Barger:1992ac,Deshpande:1994qm,Singh:1998dh,Parida:1997wx}
\begin{eqnarray}
\frac{dh_t}{dt} &=& \frac{h_t}{16\pi^2} \left(6h_t^2+h_b^2-\sum_{i=1}^3c_i \ g_i^2\right)+ \nonumber \\
& &\frac{h_t}{(16\pi^2)^2} \left[\sum_{i=1}\left(c_ib_i+\frac{c_i^2}{2}\right)g_i^4 +g_1^2g_2^2+\frac{136}{45}g_1^2g_3^2+ 8g_2^2g_3^2\right.+ \nonumber \\
& & \left. \left(\frac{6}{5}g_1^2+6g_2^2+16g_3^2\right)h_t^2 +\frac{2}{5}g_1^2h_b^2-22h_t^4-5h_b^4-5h_t^2h_b^2-h_b^2h_{\tau}^2\right], \nonumber \\
\nonumber \\
\frac{dh_b}{dt} &=& \frac{h_b}{16\pi^2} \left(6h_b^2+h_{\tau}^2+h_t^2-\sum_{i=1}^3 c_i^{'} \ g_i^2 \right)+ \nonumber \\
& &\frac{h_b}{(16\pi^2)^2}\left[\sum_{i=1}\left(c'_ib_i+\frac{c_i^{'2}}{2}\right)g_i^4+g_1^2g_2^2+\frac{8}{9}g_1^2g_3^2+8g_2^2g_3^2+\left( \frac{2}{5}g_1^2+6g_2^2+16g_3^2\right)\right. h_b^2 \nonumber \\
& & \left. +\frac{4}{5}g_1^2h_t^2+\frac{6}{5}g_1^2h_{\tau}^2-22h_b^4-3h_{\tau}^4-5h_t^4-5h_b^2h_t^2-3h_b^2h_{\tau}^2 \right], \nonumber \\
\nonumber \\
\frac{dh_{\tau}}{dt} &=& \frac{h_{\tau}}{16\pi^2} \left(4h_{\tau}^2+3h_b^2-\sum_{i=1}^3c_i^{''} \ g_i^2\right) \nonumber \\
& & +\frac{h_{\tau}}{(16\pi^2)^2}\left[\sum_{i=1}\left(c''_ib_i+\frac{c_i^{''2}}{2}\right)g_i^4+\frac{9}{5}g_1^2g_2^2+\left(\frac{6}{5}g_1^2+6g_2^2\right)h_{\tau}^2 \right. \nonumber \\
& & +\left.\left(\frac{-2}{5}g_1^2+16g_3^2\right)h_b^2 + 9h_b^4-10h_{\tau}^4-3h_b^2h_t^2-9h_b^2h_{\tau}^2 \right], \label{yrges}
\end{eqnarray}
where

\[
	\begin{matrix}
	c_i =
	\end{matrix}
	\begin{pmatrix}
	\frac{13}{15}, 3, \frac{16}{13} 
	\end{pmatrix}
	\begin{matrix}
	, \hspace{0.3cm} c_{i}^{'} =
	\end{matrix}
	\begin{pmatrix}
	\frac{7}{15}, 3, \frac{16}{3}  
	\end{pmatrix}
	\begin{matrix}
	, \hspace{0.3cm} c_{i}^{''} =
	\end{matrix}
	\begin{pmatrix}
	\frac{9}{5}, 3, 	0  
	\end{pmatrix}.
	\]

Yukawa RGEs for non-supersymmetric case,
\begin{eqnarray*}
\frac{dh_t}{dt} &=& \frac{h_t}{16\pi^2} \left(\frac{3}{2} h_t^2-\frac{3}{2}h_b^2+Y_2(S)-\sum_{i=1}^3c_i \ g_i^2\right)+ \nonumber \\
& & \frac{h_t}{(16\pi^2)^2} \left[\frac{1187}{600}g_1^4-\frac{23}{4}g_2^4-108g_3^4-\frac{9}{20}g_1^2g_2^2+\frac{19}{15}g_1^2g_3^2+9g_2^2g_3^2 \right. \nonumber \\
& & +\left. \left(\frac{223}{80}g_1^2+\frac{135}{16}g_2^2+16g_3^2\right)h_t^2-\left(\frac{43}{80}g_1^2-\frac{9}{16}g_2^2+16g_3^2 \right)h_b^2\right. \nonumber \\
& & +\left. \frac{5}{2}Y_4(S)-2\lambda\left(3h_t^2+h_b^2 \right)+\frac{3}{2}h_t^4-\frac{5}{4}h_t^2h_b^2+\frac{11}{4}h_b^4\right. \nonumber \\
& & +\left. Y_2(S)\left(\frac{5}{4}h_b^2-\frac{9}{4}h_t^2\right)- \chi_4(S)+\frac{3}{2}\lambda^2\right], \nonumber \\
\nonumber \\
\end{eqnarray*}

\begin{eqnarray*}
\frac{dh_b}{dt} &=& \frac{h_b}{16\pi^2} \left(\frac{3}{2}h_b^2-\frac{3}{2}h_t^2+Y_2(S)-\sum_{i=1}^3 c_i^{'} \ g_i^2 \right)+ \nonumber \\
& &\frac{h_b}{(16\pi^2)^2} \left[-\frac{127}{600}g_1^4-\frac{23}{4}g_2^4-108g_3^4-\frac{27}{20}g_1^2g_2^2+\frac{31}{15}g_1^2g_3^2+9g_2^2g_3^2\right. \nonumber  \\
& & \left. -\left(\frac{79}{80}g_1^2-\frac{9}{16}g_2^2+16g_3^2\right)h_t^2+\left(\frac{187}{80}g_1^2+\frac{135}{16}g_2^2+16g_3^2 \right)h_b^2 \right. \nonumber \\
& & \left. +\frac{5}{2}Y_4(S)-2\lambda\left(h_t^2+3h_b^2 \right)+\frac{3}{2}h_b^4-\frac{5}{4}h_t^2h_b^2+\frac{11}{4}h_t^4\right. \nonumber \\
& &\left. +Y_2(S)\left(\frac{5}{4}h_t^2-\frac{9}{4}h_b^2\right)-\chi_4(S)+\frac{3}{2}\lambda^2\right], \nonumber \\
\nonumber \\
\end{eqnarray*}

\begin{eqnarray}
\frac{dh_{\tau}}{dt} &=& \frac{h_{\tau}}{16\pi^2} \left(\frac{3}{2}h_{\tau}^2+Y_2(S)-\sum_{i=1}^3c_i^{''} \ g_i^2\right)+ \nonumber \\
& &  \frac{h_{\tau}}{(16\pi^2)^2} \left[\frac{1371}{200}g_1^4-\frac{23}{4}g_2^4-\frac{27}{20}g_1^2g_2^2+ \left(\frac{387}{80}g_1^2+\frac{135}{16}g_2^2\right)h_{\tau}^2 +\frac{5}{2}Y_4(S) \right. \nonumber \\
& & \left. -6\lambda h_t^2+\frac{3}{2}h_{\tau}^4 -\frac{9}{4}Y_2(S)h_{\tau}^2-\chi_4(S)+\frac{3}{2}\lambda^2 \right], \nonumber \\
\nonumber \\
\end{eqnarray}

\begin{eqnarray}
\frac{d\lambda }{dt} &=& \frac{1}{16\pi^2} \left[\frac{9}{4}\left(\frac{3}{25}g_1^4+\frac{2}{5}g_1^2g_2^2+g_2^4\right)-\left(\frac{9}{5}g_1^2+9g_2^2\right)\lambda+4Y_2(S)\lambda-4H(S)+12\lambda^2\right]+ \nonumber \\
& & \frac{1}{(16\pi^2)^2} \left[-78\lambda^3+18\left(\frac{3}{5}g_1^2+3g_2^2\right)\lambda^2+\left(-\frac{73}{8}g_2^4+\frac{117}{20}g_1^2g_2^2+\frac{1887}{200}g_1^4\right)\lambda\right.\nonumber \\
& & \left.+\frac{305}{8}g_2^6-\frac{867}{120}g_1^2g_2^4-\frac{1677}{200}g_1^4g_2^2-\frac{3411}{1000}g_1^6-64g_3^2\left(h_t^4+h_b^4\right)\right. \nonumber\\
& &\left. -\frac{8}{5}g_1^2\left(2h_t^4-h_b^4+3h_{\tau}^4\right)-\frac{3}{2}g_2^4Y_2(S)+10\lambda Y_4(S)+
\frac{3}{5}g_1^2\left(-\frac{57}{10}g_1^2+21g_2^2\right)h_t^2 \right. \nonumber \\
& & \left. +\left(\frac{3}{2}g_1^2+9g_2^2\right)h_b^2+\left(-\frac{15}{2}g_1^2+11g_2^2\right)h_{\tau}^2
-24\lambda^2Y_2(S)-\lambda H(S)+6\lambda h_t^2h_b^2 \right. \nonumber \\ 
& & \left. +20\left(3h_t^6+3h_b^6+h_{\tau}^6\right)-12\left(h_t^4h_b^2+h_t^2h_b^4\right)\right], \label{yrgens}
\end{eqnarray}

where
\begin{eqnarray}
Y_2(S)&=&3h_t^2+3h_b^2+h_{\tau}^2,  \nonumber \\ 
Y_4(S)&=&\frac{1}{3}\left[3\sum c_ig_i^2h_t^2+3\sum c^{'}_ig_i^2h_b^2+3\sum c^{''}_ig_i^2h_{\tau}^2\right], \nonumber \\ 
\chi_4(S)&=&\frac{9}{4}\left[3h_t^4+3h_b^4+h_{\tau}^4-\frac{2}{3}h_t^2h_b^2\right], \nonumber \\
H(S)&=&3h_t^4+3h_t^4+h_{\tau}^4, \nonumber \\
\lambda &=& \frac{m_h^2}{V^2}, is \ the \ Higgs \ self \ coupling \ (m_h= \ Higgs \ mass ).
\end{eqnarray}

with the values of beta function coefficients for non-SUSY case and \\
	\[
	\begin{matrix}
	c_i =
	\end{matrix}
	\begin{pmatrix}
	0.85, 2.25, 	8.00 
	\end{pmatrix}
	\begin{matrix}
	, \hspace{0.3cm} c_{i}^{'} =
	\end{matrix}
	\begin{pmatrix}
	0.25, 2.25, 	8.00  
	\end{pmatrix}
	\begin{matrix}
	, \hspace{0.3cm} c_{i}^{''} =
	\end{matrix}
	\begin{pmatrix}
	2.25, 	2.25, 0.00  
	\end{pmatrix}.
	\]

\clearpage

\section*{Tables}  
\begin{table}[t]
\centering
\begin{tabular}{c|llllllllll}
\hline
\multirow{2}{*}{}   &\multicolumn{9}{c} {}\\
{$(\psi_{1})_{0}/^{\circ}\rightarrow$} & $0.0$ & $45$ & $90$  & $135$ & $180$ & $225$ & $270$ & $315$ & $360$ \\
\hline
\hline

      $(\theta_{23})_{0}/^{\circ}\rightarrow$   & 37.240 &  37.240   &   37.240   &  37.240  & 37.240 &   37.240  & 37.240 &   37.240 &   37.240 \\ 
      $(\theta_{12})_{0}/^{\circ}\rightarrow$   & 29.160 &  29.160   &   29.160   &  29.160  & 29.160 &   29.160  & 29.160 &   29.160 &   29.160 \\
      $(\theta_{13})_{0}/^{\circ}\rightarrow$   & 7.974  &  7.974    &   7.974    &  7.974   & 7.974  &   7.974   & 7.974  &   7.974  &   7.974  \\
      $(m_1)_{0}\times 10^{-2}eV\rightarrow$   & 1.370 &  1.375   &   1.380   &  1.390  & 1.395 &   1.389 &  1.380  &  1.375 &  1.370 \\
      $(m_3)_{0}\times 10^{-2}eV\rightarrow$   & 7.791 &  7.802   &   7.626   &  7.460  & 7.470 &   7.469 &  7.522  &  7.801 &  7.779 \\

 \hline
\end{tabular} 
\caption{\footnotesize The initial values of the neutrino parameters ($(\theta_{ij})_{0}$ and $(m_{i})_{0}$) against varying Majorana phase ($(\psi_{1})_{0}$, $(\psi_{2})_{0}$,with, $(\psi_{1})_{0}=(\psi_{2})_{0}$). The $m_s$ and the SS scale are fixed at $5\, TeV$ and $10^{14}\,GeV$ respectively. We choose, the initial value of the Dirac phase, $(\delta)_0=90^{\circ}$, and $(m_2)_0=2.340\times 10^{-2}eV$. The purpose of this study is to achieve the numerical values of the parameters within $3\sigma$ range at EW scale. }
\label{vmp}
\end{table}

\begin{table}[t]
\centering
\renewcommand{\arraystretch}{1.0}
\begin{tabular}{c|ccccccc|ccccccccc}
\cline{1-15}
\multicolumn{1}{c|}{Input} & \multicolumn{14}{c}{\multirow{1}{*}{Different possible sets of neutrino parameters input values}} \\ 
\multicolumn{1}{c|}{$\nu$ Para-} & \multicolumn{14}{c}{\multirow{1}{*}{at the seesaw scale}} \\ 
\multicolumn{1}{c|}{meters}  & \tiny$\downarrow$  & \tiny$\downarrow$ & \tiny$\downarrow$ & \tiny$\downarrow$ & \tiny$\downarrow$ & \tiny$\downarrow$ & \tiny$\downarrow$ & \tiny$\downarrow$  & \tiny$\downarrow$ & \tiny$\downarrow$ & \tiny$\downarrow$ & \tiny$\downarrow$ & \tiny$\downarrow$ & \tiny$\downarrow$  \\ \cline{2-15}
\multicolumn{1}{l|}{} & $A1$ &  $A3$ & $A5$ & $A7$ & $A9$ & $A11$ & $A13$ & $B1$ &  $B3$ & $B5$ & $B7$ & $B9$ & $B11$ & $B13$  \\ 
\cline{1-15}
$(m_1)_{0}$   &	1.51	&	1.36	&	1.32	&	1.29	&	1.27	&	1.25	&	1.23 &	1.51		&	1.38	&	1.33	&	1.30	&	1.27	&	1.25	&	1.24  \\
$(m_2)_0$	&	2.34	&	2.34	&	2.34	&	2.34	&	2.34	&	2.34 	&	2.34 &	2.34		&	2.34	&	2.34	&	2.34	&	2.34	&	2.34 	&	2.34  \\
$(m_3)_0$   &	7.61	&	7.74	&	7.79	&	7.81	&	7.86	&	7.92	&	7.92 &	7.46	    &	7.56	&	7.58	&	7.63	&	7.65	&	7.68	&	7.72  \\
\hline 
$(\theta_{12})_0/^{\circ}$ &	30.36 &	31.05 &	31.45 &	31.46 &	31.51	&	31.62 	&	31.79  &	30.36	&	31.22	&	31.79	&	31.79	&	31.79	&	31.79 	&	31.79  \\
$(\theta_{13})_0/^{\circ}$ &	8.42  &	8.53  &	8.53  &	8.53  &	8.53	&	8.53	&	8.53   &	8.93	&	9.05	&	9.05	&	9.05	&	9.05	&	9.05	&	9.05  \\
$(\theta_{23})_0/^{\circ}$ &	37.12 &	37.61 & 37.99 &	37.99 &	38.04	&	38.15 	&	38.31  &	38.12	&	38.26	&   38.80	&	38.80	&	38.80	&	38.80 	&	38.80  \\
$q$ 				    &	0.96  &	0.95  &	0.95  &	0.95  &	0.95	&	0.95 	&	0.95   &	0.97	&	0.95	&	0.95	&	0.95	&	0.95   &	0.95 	&	0.95  \\
\hline 
$(\psi_1)_0/^{\circ}$ &	45.0 &	45.0 &	45.0 &	45.0 &	45.0 &	45.0 &	45.0  	&	45.0    &	45.0    &	45.0     & 45.0    & 45.0      &	45.0    &	45.0  \\
$(\delta)_{0}/^{\circ}$			     &	90.0 &	90.0 &	90.0 &	90.0 &	90.0 &	90.0 &	90.0    &	270.0	&	270.0   &   270.0	 & 270.0	& 270.0	    &	270.0	&	270.0  \\ 
\hline
\end{tabular} 
\caption{\footnotesize 
The Table for different sets of input parameters to be used for subsequent analysis.The $(\theta_{23})_{0}$ is connected to $(\theta_{13})_0$ and $(\theta_{12})_{0}$ via the S.C relation as presumed in eq.\,(\ref{comp2}). We choose only the $(\theta_{13})_0$ and $(\theta_{12})_0$ as input. The Majorana phase $(\psi_{1})_0$ is fixed at $45^{\circ}$. The sets $A1$, $A3$,...$A{13}$ represent the collection of initial inputs to be attributed to the parameters at the SS scale, for the $m_{s}$ scale being fixed at $1$, $3$...$13$\,TeV respectively, with $(\delta)_{0}=90^{\circ}$. The SS scale is fixed at $10^{14}$\,GeV. The sets $B1$, $B2$,... $B13$ are similar to the sets $A{1}$, $A3$...$A13$, respectively, except for the former, $(\delta)_0=270^0$. The numerical entries are so adjusted for a specific $m_{s}$ scale (say, $A5$ at $5TeV$) so that after running the RGEs, the parameters at the EW scale lie within the $3\sigma$ range.}
\label{90i}
\end{table}

\begin{table}[t]
\centering
\renewcommand{\arraystretch}{1.0}
\begin{tabular}{c|ccccccc|ccccccccc}
\cline{1-15}
\multicolumn{1}{c|}{$m_s$} & \multicolumn{14}{c}{\multirow{1}{*}{$\Delta m_{12}^2 \ (\times 10^{-5} eV^2)$}} \\ 
\multicolumn{1}{c|}{in} & \multicolumn{14}{c}{\multirow{1}{*}{at EW scale }} \\ 
 \cline{2-15}
\multicolumn{1}{l|}{TeV} & $A1$ &  $A3$ & $A5$ & $A7$ & $A9$ & $A11$ & $A13$ & $B1$ &  $B3$ & $B5$ & $B7$ & $B9$ & $B11$ & $B13$  \\ 
\cline{1-15}
      1.0   &\underline{ \textbf{7.56}}   &  11.11  &  13.00    & 12.98  & 13.50    &  13.86   &  14.19  & \underline{\textbf{7.56}}     & 11.08    & 12.34    & 13.01  & 13.52  & 13.89  &  14.21  \\
      3.0   & 1.79   &  \underline{\textbf{7.56}} &  9.16    & 9.82    & 10.81    &  10.89   &  11.25  & 1.88    & \underline{\textbf{7.57}}     & 9.05     & 9.63   & 10.42  & 10.84  & 11.20   \\
      5.0   & $\times$&  5.94  &  \underline{\textbf{7.57}}   & 8.42    & 9.06    &  9.47   &  9.844  & $\times$ & 5.95     & \underline{\textbf{7.58}}     & 8.43   & 9.06   & 9.50   & 9.88   \\
      7.00   & $\times$&  4.87 &  6.64   & \underline{\textbf{7.55}}    & 8.21    &  8.64   &  9.01  & $\times$ & 4.87     & 6.64     & \textbf{7.55}  & 8.20   & 8.66   & 9.06    \\
      9.0   & $\times$&  4.01  &  5.93    & 6.89    & \underline{\textbf{7.57} }  &  \underline{8.01}   &  8.400  & $\times$ & 4.00     & 5.93     & 6.88   & \underline{\textbf{7.56}}  & \underline{8.03}   & 8.44   \\
      11.0  & $\times$&  3.30  &  5.39    & 6.38   & \underline{7.08}  & \underline{\textbf{ 7.54}}   &  \underline{7.92}  & $\times$ & 3.27     & 5.37     & 6.36   & \underline{7.07}   & \underline{\textbf{7.56}}   & \underline{7.97}   \\
      13.0  & $\times$&  2.64  &  4.919    & 5.95    & 6.675    &  \underline{7.13}   &  \underline{\textbf{7.53}}  & $\times$ & 2.61     & 4.89     & 5.29   & 6.65   & \underline{7.15}   & \underline{\textbf{7.55}}  \\  
  
 \hline
\end{tabular} 
\caption{\footnotesize The fluctuations of the solar mass squared difference after RG evolution, at the EW scale have been studied, against changing $m_{s}$ at a constant SS scale. The $Aj$ or $Bj$ correspond to the set of initial entries at a constant $m_{s}$ as mentioned in Table.\,(\ref{90i}). The diagonal entries marked in bold text reflect the output values of, $\Delta m_{21}^2$ within $3\sigma$ for which the initial entries of $Aj$ or $Bj$ were tuned at constant $m_s$. On keeping a input data set (say, $A5$) fixed, if the $m_s$ scale is varied, one sees that, against the radiative correction, the value of $\Delta m_{21}^2$ at the EW scale fluctuates. If $m_s$ is lesser, the fluctuation is more. The output values which lies within $3\sigma$ are underlined. The irrelevant output are omitted with `$\times$' sign.}
\label{90m12}
\end{table}

\begin{table}[t]
\centering
\renewcommand{\arraystretch}{1.0}
\begin{tabular}{c|ccccccc|ccccccccc}
\cline{1-15}
\multicolumn{1}{c|}{$m_s$} & \multicolumn{14}{c}{\multirow{1}{*}{$\Delta m_{23}^2\ (\times 10^{-3} eV^2)$}} \\ 
\multicolumn{1}{c|}{in} & \multicolumn{14}{c}{\multirow{1}{*}{at EW scale for different sets of inputs}} \\ 
 \cline{2-15}
\multicolumn{1}{l|}{TeV} & $A1$ &  $A3$ & $A5$ & $A7$ & $A9$ & $A11$ & $A13$ & $B1$ &  $B3$ & $B5$ & $B7$ & $B9$ & $B11$ & $B13$  \\ 
\cline{1-15}
      1.0   & \underline{\textbf{2.51}}    & \underline{2.65}    &  2.86    &  2.703  & 2.74   & 2.80   &  2.80  & \underline{\textbf{2.49}}   & \underline{2.59}     & \underline{2.62}     & \underline{2.66}   & 2.68   & 2.71   & 2.74\\
      3.0   & 2.40     &  \underline{\textbf{2.53}}    &  \underline{2.62}    &  \underline{2.57}  & \underline{2.62}   & \underline{2.67}   &  \underline{2.67}  & 2.40    & \underline{\textbf{2.50}}     & \underline{2.53}     & \underline{2.53}   & \underline{2.59}   & \underline{2.62}   & \underline{2.64}\\
      5.0   & $\times$ &  \underline{2.48} &   \underline{\textbf{2.51}}   & \underline{2.51} & \underline{2.56}   & \underline{2.61}   &  \underline{2.61}  & $\times$ & \underline{2.45}    & \underline{\textbf{2.49}}    & \underline{2.52}   & \underline{2.54}   & \underline{2.57}   & \underline{2.60}\\
      7.0   & $\times$ &  \underline{2.44}    &  \underline{2.47}    &  \underline{\textbf{2.48}}  & \underline{2.52}   & \underline{2.57}   & \underline{2.57}  & $\times$ & 2.42    & \underline{2.45}     & \underline{\textbf{2.49}}   & \underline{2.51}  & \underline{2.54}   & \underline{2.56}\\
      9.0   & $\times$ &  2.41    &  \underline{2.44}    &  \underline{2.44} & \underline{\textbf{2.49}}   & \underline{2.54}   & \underline{ 2.54}  & $\times$ & 2.40    & \underline{2.43}     & \underline{2.47}   & \underline{\textbf{2.48}}   & \underline{2.51}  & \underline{2.53}\\
      11.0  & $\times$ &  2.39    &  2.41    &  2.42  & \underline{2.46}   & \textbf{2.51}   &  \underline{2.51}  & $\times$ & 2.38   & 2.41     & \underline{2.45}   & \underline{2.46}   & \textbf{2.49}   & \underline{2.52}\\
      13.0  & $\times$ &  2.37    &  2.39    &  2.40 & \underline{2.44}   & \underline{2.49}   &  \underline{\textbf{2.44}}  & $\times$ & 2.36    & 2.39     & \underline{2.43}   & \underline{2.45}   & \underline{2.47}   & \underline{\textbf{2.49}}\\   
   
 \hline
\end{tabular} 
\caption{\footnotesize The fluctuations of an atmospheric mass squared difference after RG evolution, at the EW scale have been studied, against changing $m_{s}$, at a constant SS scale. The $Aj$ or $Bj$ correspond to the set of initial entries at a constant $m_{s}$ as mentioned in Table.\,(\ref{90i}). The diagonal entries marked in bold text reflect the output values of $\Delta m_{31}^2$ within $3\sigma$ for which the initial entries of $Aj$ or $Bj$ were tuned at a constant $m_s$. On keeping a input data set (say, $A5$) fixed, if the $m_s$ scale is varied, one sees that, against the radiative correction, the value of $\Delta m_{31}^2$ at the EW scale fluctuates. If $m_s$ is lesser, the fluctuation is more. The output values which lies within $3\sigma$ are underlined. The irrelevant results in view of $3\sigma$ bound are omitted with `$\times$' symbol.}
\label{90m23}
\end{table}

\begin{table}[t]
\centering
\renewcommand{\arraystretch}{1.0}
\begin{tabular}{c|ccccccc|ccccccccc}
\cline{1-15}
\multicolumn{1}{c|}{$m_s$} & \multicolumn{14}{c}{\multirow{1}{*}{$\theta_{23}/^{\circ}$}} \\ 
\multicolumn{1}{c|}{in} & \multicolumn{14}{c}{\multirow{1}{*}{at EW scale}} \\ 
 \cline{2-15}
\multicolumn{1}{l|}{TeV} & $A1$ &  $A3$ & $A5$ & $A7$ & $A9$ & $A11$ & $A13$ & $B1$ &  $B3$ & $B5$ & $B7$ & $B9$ & $B11$ & $B13$  \\ 
\cline{1-15}
      1.0   &\textbf{41.0}    &   41.4   &   41.6   &   41.8 &   41.8 &   41.9 &   42.1 & \textbf{41.0}    &   41.1   &   41.6   &   41.6 &   41.6 &   41.6 &   41.6\\
      3.0   & 40.6    &   \textbf{41.1}   &   41.4   &   41.5 &   41.5 &   41.6 &   41.8 & 40.7    &   \textbf{40.8}   &   41.3   &   41.3 &   41.3 &   41.3 &   41.3\\
      5.0   & $\times$ &   41.0   &   \textbf{41.3}   &   41.3 &   41.4 &   41.5 &   41.6 & $\times$ &   40.7   &   \textbf{41.2}   &   41.2 &   41.2 &   41.2 &   41.2\\
      7.0   & $\times$ &   40.9   &   41.2   &  \textbf{41.2} &   41.3 &   41.4 &   41.5 & $\times$ &   40.6   &   41.1   &   \textbf{41.1} &   41.1 &   41.1 &   41.1\\
      9.0   & $\times$ &   40.8   &   41.2   &   41.2 &   \textbf{41.2} &   41.3 &   41.4 & $\times$ &   40.5   &   41.1    &   41.1 &   \textbf{41.1}&   41.0 &   41.0\\
      11.0  & $\times$ &   40.8   &   41.1   &   41.1 &   41.1 &   \textbf{41.2}&   41.4 & $\times$ &   40.5   &   41.0   &   41.0 &   41.0 &   \textbf{41.0} &   41.0\\
      13.0  & $\times$ &   40.7   &   41.1   &   41.0 &   41.1 &   41.2 &   \textbf{41.3} & $\times$ &   40.4   &   41.0   &   41.0 &   41.0 &   41.0 &  \textbf{41.0}\\ 
  
 \hline
\end{tabular} 
\caption{\footnotesize The fluctuations of atmospheric angle  after RG evolution, at the EW scale have been studied, against changing $m_{s}$, at a constant SS scale. The $Aj$ or $Bj$ represent the set of initial entries at a constant $m_{s}$ as mentioned in Table.\,(\ref{90i}). The diagonal entries marked in bold text reflect the output values of, $\theta_{23}$ within $3\sigma$ for which the initial entries of $Aj$ or $Bj$ are  adjusted at constant $m_s$. On keeping a input data set (say, $A5$) fixed, if the $m_s$ scale is varied, one sees that, against the radiative correction, the value of $\theta_{23}$, at the EW scale fluctuates, but  a little and output values  lie within $3\sigma$ range. The irrelevant results in view of $3\sigma$ bound are omitted with `$\times$' symbol.}
\label{90t23}
\end{table}

\begin{table}[t]
\centering
\renewcommand{\arraystretch}{1.0}
\begin{tabular}{c|ccccccc|ccccccccc}
\cline{1-15}
\multicolumn{1}{c|}{$m_s$} & \multicolumn{14}{c}{\multirow{1}{*}{$\theta_{12}/^{\circ}$}} \\ 
\multicolumn{1}{c|}{in} & \multicolumn{14}{c}{\multirow{1}{*}{at EW scale}} \\ 
 \cline{2-15}
\multicolumn{1}{l|}{TeV} & $A1$ &  $A3$ & $A5$ & $A7$ & $A9$ & $A11$ & $A13$ & $B1$ &  $B3$ & $B5$ & $B7$ & $B9$ & $B11$ & $B13$  \\ 
\cline{1-15}
      1.0   & \textbf{34.6}     &   34.8   &   34.8   &   34.8 &   34.8 &   34.8 &   35.0 & \textbf{34.8}    & 35.0     & 35.3     & 35.1   & 35.1   & 35.1   & 35.0\\
      3.0   & 34.1     &  \textbf{ 34.4}   &   34.6   &   34.6 &   34.5 &   34.6 &   34.7 & 34.3    & \textbf{34.7}     & 35.0     & 34.9   & 34.9   & 34.8   & 34.7\\
      5.0   & $\times$  &   34.3   &   \textbf{34.5}   &   34.4 &   34.4 &   34.5 &   34.6 & $\times$ & 34.5     & \textbf{34.9}     & 34.7   & 34.7   & 34.7   & 34.6\\
      7.0   & $\times$  &   34.2   &   34.4   &  \textbf{34.3} &   34.3 &   34.4 &   34.5 & $\times$ & 34.4     & 34.8     &  \textbf{34.6}   & 34.7   & 34.6   & 34.5\\
      9.0   & $\times$  &   34.1   &   34.4   &   34.3 &    \textbf{34.2} &   34.3 &   34.5 & $\times$ & 34.3     & 34.7     & 34.6   &  \textbf{34.6}   & 34.5   & 34.5\\
     11.0   & $\times$  &   34.0   &   34.3   &   34.2 &   34.2 &    \textbf{34.3} &   34.4 & $\times$ & 34.3     & 34.6     & 34.5   & 34.5   &  \textbf{34.5}   & 34.48\\
     13.0   & $\times$  &   34.0   &   34.3   &   34.2 &   34.2 &   34.2 &   \textbf{34.4} & $\times$ & 34.2     & 34.6     & 34.5   & 34.5   & 34.4   &  \textbf{34.4}\\ 
     
 \hline
\end{tabular} 
\caption{\footnotesize The fluctuation of solar angle  after RG evolution, at the EW scale is studied, against changing $m_{s}$, at a constant SS scale. The $Aj$ or $Bj$ represent the set of initial entries at a constant $m_{s}$ as mentioned in Table.\,(\ref{90i}). The diagonal entries marked in bold texts reflect the output values of, $\theta_{12}$ within $3\sigma$ for which the initial entries of $Aj$ or $Bj$ are  adjusted at constant $m_s$. On keeping a input data set (say, $A5$) fixed, if the $m_s$ scale is varied, one sees that, against the radiative correction, the value of $\theta_{12}$, at the EW scale fluctuates, but  the variations are a little and the output values  lie within $3\sigma$ range.}
\label{90t12}
\end{table}

\begin{table}[t]
\centering
\renewcommand{\arraystretch}{1.0}
\begin{tabular}{c|ccccccc|ccccccccc}
\cline{1-15}
\multicolumn{1}{c|}{$m_s$} & \multicolumn{14}{c}{\multirow{1}{*}{$\theta_{13}/^{\circ}$}} \\ 
\multicolumn{1}{c|}{in} & \multicolumn{14}{c}{\multirow{1}{*}{at EW }} \\ 
 \cline{2-15}
\multicolumn{1}{l|}{TeV} & $A1$ &  $A3$ & $A5$ & $A7$ & $A9$ & $A11$ & $A13$ & $B1$ &  $B3$ & $B5$ & $B7$ & $B9$ & $B11$ & $B13$  \\ 
\cline{1-15}
      1.0   & \textbf{8.4}      &   8.6  &   8.6   &   8.6 &   8.6 &   8.6 &   8.6 & \textbf{8.4}          &  8.5    & 8.4     & 8.4   & 8.4   & 8.4   & 8.4\\
      3.0   & 8.4      &   \textbf{8.5}   &   8.5   &   8.5 &   8.5 &   8.5 &   8.5 & 8.3          &  \textbf{8.4}    & 8.4     & 8.3   & 8.4   & 8.4   & 8.4\\
      5.0   & $\times$  &   8.4 &   \textbf{8.4}   &   8.5 &   8.50 &   8.50 &   8.5 & $\times$      &  8.4    & \textbf{8.3}     & 8.3   & 8.3   & 8.3   & 8.3\\
      7.0   & $\times$  &   8.4   &   8.4   &   \textbf{8.4} &   8.4 &   8.4 &   8.4 & $\times$      &  8.3    & 8.3     & \textbf{8.3}   & 8.3   & 8.3   & 8.3\\
      9.0   & $\times$  &   8.4 &   8.4  &   8.4 &   \textbf{8.4} &   8.4 &   8.4 & $\times$      &  8.3    & 8.3     & 8.3   & \textbf{8.3}   & 8.3  & 8.3\\
     11.0   & $\times$  &   8.4   &   8.4   &   8.4 &   8.4 &   \textbf{8.4} &   8.4 & $\times$      &  8.3    & 8.3     & 8.3   & 8.3   & \textbf{8.3}   & 8.3\\
     13.0   & $\times$  &   8.4   &   8.4   &   8.4 &   8.4 &   8.4 &   \textbf{8.4} & $\times$      &  8.3    & 8.3    & 8.3   & 8.3   & 8.3   & \textbf{8.3}\\ 
 
 \hline
\end{tabular} 
\caption{\footnotesize The fluctuation of the reactor angle after RG evolution, at the EW scale is investigated , against changing $m_s$, at a constant SS scale. The $Aj$ or $Bj$ represent the set of initial entries at a constant $m_{s}$ as mentioned in Table.\,(\ref{90i}). The diagonal entries marked in Bold texts represent the output values of, $\theta_{23}$ within $3\sigma$ for which the initial entries of $Aj$ or $Bj$ are  adjusted at a constant $m_s$. On keeping an input data set (say, $A5$) fixed, if the $m_s$ scale is varied, one sees that, against the radiative correction, the value of $\theta_{23}$ at the EW scale fluctuates. The fluctuation is very feeble against the varying $m_s$. The irrelevant results in view of $3\sigma$ bound are omitted with `$\times$' symbol.}
\label{90t13}
\end{table}

\begin{table}[t]
\centering
\renewcommand{\arraystretch}{1.0}
\begin{tabular}{c|ccccccc|ccccccccc}
\cline{1-15}
\multicolumn{1}{c|}{$m_s$} & \multicolumn{14}{c}{\multirow{1}{*}{$m_1\times 10^{-3}eV$}} \\ 
\multicolumn{1}{c|}{in} & \multicolumn{14}{c}{\multirow{1}{*}{at EW scale for different sets of inputs}} \\ 
 \cline{2-15}
\multicolumn{1}{l|}{TeV} & $A1$ &  $A3$ & $A5$ & $A7$ & $A9$ & $A11$ & $A13$ & $B1$ &  $B3$ & $B5$ & $B7$ & $B9$ & $B11$ & $B13$  \\ 
\cline{1-15}
      1.0   & 10.30    &   9.43   &   9.35   &   8.83 &   8.66 &   8.53 &   8.40 & 10.29         &  9.46    &  9.04    & 8.83   & 8.67   & 8.55   & 8.44\\
      3.0   & 9.63     &   8.82   &   8.56   &   8.26 &   8.10 &   7.98 &   7.85 & 9.62          &  8.85    &  8.45    & 8.06   & 8.11   & 8.00   & 7.89\\
      5.0   & $\times$ &   8.53   &   8.17   &   7.99 &   7.83 &   7.71 &   7.59 & $\times$      &  8.56    &  8.17    & 7.99   & 7.84   & 7.73   & 7.63\\
      7.0   & $\times$ &   8.34   &   7.99   &   7.81 &   7.65 &   7.54 &   7.42 & $\times$      &  8.36    &  7.99    & 7.81   & 7.66   & 7.56   & 7.46\\
      9.0   & $\times$ &   8.19   &   7.84   &   7.67 &   7.52 &   7.40 &   7.29 & $\times$      &  8.21    &  7.85    & 7.67   & 7.53   & 7.42   & 7.33\\ 
      11.0  & $\times$ &   8.07   &   7.73   &   7.56 &   7.41 &   7.30 &   7.18 & $\times$      &  8.10    &  7.73    & 7.56   & 7.42   & 7.31   & 7.22\\ 
      13.0  & $\times$ &   7.97   &   7.63   &   7.46 &   7.31 &   7.20 &   7.09 & $\times$      &  7.99    &  7.63    & 7.46   & 7.32   & 7.22   & 7.11\\ 
      
\hline
\end{tabular} 
\caption{\footnotesize The fluctuations of $m_1$ after RG evolution at the EW scale have been studied, against changing $m_{s}$, at a constant SS scale. The $Aj$ or $Bj$ correspond to the set of initial entries at a constant $m_{s}$ as mentioned in Table.\,(\ref{90i}). On keeping an input data set (say, $A5$) fixed, if the $m_s$ scale is varied, one sees that, against the radiative correction, the value of $m_1$ at EW scale fluctuates. The irrelevant results in view of $3\sigma$ bound are omitted with `$\times$' symbol.}
\label{90m1}
\end{table}

\begin{table}[t]
\centering
\renewcommand{\arraystretch}{1.0}
\begin{tabular}{c|ccccccc|ccccccccc}
\cline{1-15}
\multicolumn{1}{c|}{$m_s$} & \multicolumn{14}{c}{\multirow{1}{*}{$m_2\times 10^{-2}eV$}} \\ 
\multicolumn{1}{c|}{in} & \multicolumn{14}{c}{\multirow{1}{*}{at EW scale for different sets of inputs}} \\ 
 \cline{2-15}
\multicolumn{1}{l|}{TeV} & $A1$ &  $A3$ & $A5$ & $A7$ & $A9$ & $A11$ & $A13$ & $B1$ &  $B3$ & $B5$ & $B7$ & $B9$ & $B11$ & $B13$  \\ 
\cline{1-15}
      1.0   & 1.34     &   1.41   &   1.47   &   1.44 &   1.44 &   1.45 &   1.45 &   13.47       &   1.42   &  1.43    & 1.44   & 1.45   & 1.46   & 1.46\\
      3.0   & 1.05     &   1.23   &   1.28   &   1.29 &   1.30 &   1.31 &   1.31 &   10.55       &   1.24   &  1.27    & 1.27   & 1.30   & 1.31   & 1.32\\
      5.0   & $\times$ &   1.15   &   1.19   &   1.21 &   1.23 &   1.24 &   1.24 &$\times$       &   1.15   &  1.19    & 1.22   & 1.23   & 1.24   & 1.25\\
      7.0   & $\times$ &   1.08   &   1.14   &   1.16 &   1.18 &   1.19 &   1.20 & $\times$      &   1.09   &  1.14    & 1.17   & 1.19   & 1.20   & 1.21\\
      9.0   & $\times$ &   1.03   &   1.09   &   1.13 &   1.15 &   1.16 &   1.17 &$\times$       &   1.04   &  1.09    & 1.13   & 1.15   & 1.16   & 1.18\\
      11.0  & $\times$ &   0.99   &   1.06   &   1.10 &   1.12 &   1.13 &   1.14 &$\times$       &   0.99   &  1.07    & 1.10   & 1.12   & 1.14   & 1.15\\
      13.0  & $\times$ &   0.95   &   1.03   &   1.07 &   1.09 &   1.11 &   1.12 &$\times$       &   0.95   &  1.04    & 1.07   & 1.10   & 1.11   & 1.12\\ 
       
\hline
\end{tabular} 
\caption{\footnotesize The fluctuations of $m_2$ after RG evolution at the EW scale is studied, against changing $m_{s}$, at a constant SS scale. The $Aj$ or $Bj$ correspond to the set of initial entries at a constant $m_{s}$ as mentioned in Table.\,(\ref{90i}). On keeping an input data set (say, $A5$) fixed, if the $m_s$ scale is varied, one sees that, against the radiative correction, the value of $m_2$, at the EW scale fluctuates. The irrelevant results in view of $3\sigma$ bound are omitted with `$\times$' symbol.}
\label{90m2}
\end{table}

\begin{table}[t]
\centering
\renewcommand{\arraystretch}{1.0}
\begin{tabular}{c|ccccccc|ccccccccc}
\cline{1-15}
\multicolumn{1}{c|}{$m_s$} & \multicolumn{14}{c}{\multirow{1}{*}{$m_3\times 10^{-2}eV$}} \\ 
\multicolumn{1}{c|}{in} & \multicolumn{14}{c}{\multirow{1}{*}{at EW scale for different sets of inputs}} \\ 
 \cline{2-15}
\multicolumn{1}{l|}{TeV} & $A1$ &  $A3$ & $A5$ & $A7$ & $A9$ & $A11$ & $A13$ & $B1$ &  $B3$ & $B5$ & $B7$ & $B9$ & $B11$ & $B13$  \\ 
\cline{1-15}
      1.0   & 5.12     &   5.24   &   5.43   &   5.27 &   5.31 &   5.36 &   5.36 & 5.09          &  5.17    & 5.20     & 5.23   & 5.25   & 5.27   & 5.30 \\
      3.0   & 4.99     &   5.11   &   5.19   &   5.14 &   5.18 &   5.23 &   5.23 & 5.00          &  5.08    & 5.10     & 5.10   & 5.15   & 5.18   & 5.20\\
      5.0   & $\times$ &   5.05   &   5.07   &   5.08 &   5.12 &   5.17 &   5.17 & $\times$      &  5.03    & 5.05     & 5.09   & 5.10   & 5.13   & 5.15\\
      7.0   & $\times$ &   5.01   &   5.03   &   5.04 &   5.08 &   5.12 &   5.12 & $\times$      &  4.99    & 5.02     & 5.05   & 5.07   & 5.09   & 5.12\\
      9.0   & $\times$ &   4.98   &   5.00   &   5.00 &   5.04 &   5.09 &   5.09 & $\times$      &  4.96    & 4.99     & 5.02   & 5.04   & 5.07   & 5.09\\
      11.0  & $\times$ &   4.95   &   4.97   &   4.98 &   5.02 &   5.06 &   5.06 & $\times$      &  4.94    & 4.97     & 5.00   & 5.02   & 5.04   & 5.07\\
      13.0  & $\times$ &   4.93   &   4.95   &   4.95 &   4.99 &   5.04 &   5.04 & $\times$      &  4.92    & 4.95     & 4.98   & 5.00   & 5.02   & 5.04\\ 
 \hline
\end{tabular} 
\caption{\footnotesize The fluctuations of $m_3$ after RG evolution, at the EW scale is studied, against changing $m_{s}$, at a constant SS scale. The $Aj$ or $Bj$ correspond to the set of initial entries at a constant $m_{s}$ as mentioned in Table.\,(\ref{90i}). On keeping an input data set (say, $A5$) fixed, if the $m_s$ scale is varied, one sees that, against the radiative correction, the value of $m_3$, at the EW scale fluctuates. The irrelevant results in view of $3\sigma$ bound are omitted with `$\times$' symbol.}
\label{90m3}
\end{table}

\begin{table}[t]
\centering
\renewcommand{\arraystretch}{1.0}
\begin{tabular}{c|ccc|cc|cc}
\cline{1-8}
\multicolumn{1}{c|}{Variation} & \multicolumn{7}{c}{\multirow{1}{*}{Effect}} \\ 
\multicolumn{1}{c|}{of} & \multicolumn{7}{c}{\multirow{1}{*}{of varying $m_s$ and SS }} \\ 
\multicolumn{1}{c|}{$m_s$ and SS} & \multicolumn{7}{c}{{on the neutrino parameters}} \\ \cline{2-8}
\multicolumn{1}{c|}{scale} & $\theta_{12}$ &  $\theta_{13}$ & $\theta_{23}$ & $\Delta m_{21}^2$ & $\Delta m_{31}^2$ &  $\delta$ & $\psi_1$ \\ 
\cline{1-8}
Increasing $m_s$ $\rightarrow$  &	$-$	&	$-$	&	$-$	&	$-$	&	$-$	&	$-$	&	$+$ \\
Decreasing $m_s$ $\rightarrow$  &	$+$	&	$+$	&	$+$	&	$+$	&	$+$ &	$+$ &   $-$ \\
\hline 
Increasing SS $\rightarrow$ &	$+$	&	$-$	&	$+$	&	$-$	&	$-$	&	$+$	&	$-$ \\
Decreasing SS $\rightarrow$ &	$-$	&	$+$	&	$-$	&	$+$	&	$+$	&	$-$ &   $+$ \\
\hline
\end{tabular} 
\caption{\footnotesize Here we show the different effects each neutrino parameters receive due to the variation of $m_s$ and SS. An increase in $m_s$ causes a negative effect on all the EW scale neutrino parameters values, except for the Majorana phases (for decreasing $m_s$ the finding is reverse), whereas variation in SS has unequal effects (positive effect on some parameters and negative effects on other parameters). The `$-$' sign indicates the negative effect, whereas the `$+$' sign indicate the positive contribution due to varying $m_s$ and SS.}
\label{eff}
\end{table} 
\clearpage

\section*{Graphs}

\section*{Acknowledgements}
One of the author (K.S. Singh) would like to thank UGC for their financial support (grant no. F.7-65/2007(BSR)).

\clearpage
\bibliography{ps}

\end{document}